\documentclass[aps,twocolumn,amsmath,amssymb,floatfix,superscriptaddress,nofootinbib]{revtex4-1}

\usepackage{graphicx}
\usepackage{dcolumn}
\usepackage{bm}
\usepackage{latexsym}
\usepackage{color}
\usepackage[colorlinks=true,linkcolor=blue,citecolor=blue]{hyperref}
\usepackage{xcolor}

\begin{document}

\title{Physics of relativistic collisionless shocks:\\ II Dynamics of the background plasma}

\author{Martin Lemoine}
\affiliation{Institut d'Astrophysique de Paris, CNRS -- Sorbonne Universit\'e, 98 bis boulevard Arago, F-75014 Paris}
\author{Arno Vanthieghem}
\affiliation{Institut d'Astrophysique de Paris, CNRS -- Sorbonne Universit\'e, 98 bis boulevard Arago, F-75014 Paris}
\affiliation{Sorbonne Universit\'e, Institut Lagrange de Paris (ILP),
98 bis bvd Arago, F-75014 Paris, France}
\author{Guy Pelletier} 
\affiliation{UJF-Grenoble, CNRS-INSU, Institut de Plan\'etologie et d'Astrophysique de
Grenoble (IPAG), F-38041 Grenoble, France}
\author{Laurent Gremillet}
\affiliation{CEA, DAM, DIF, F-91297 Arpajon, France}

\date{\today}

\begin{abstract}  
In this second paper of a series, we discuss the dynamics of a plasma entering the precursor of an unmagnetized, relativistic collisionless pair shock. We discuss how this background plasma is decelerated and heated through its interaction with a microturbulence that results from the growth of a current filamentation instability (CFI) in the shock precursor. We make use, in particular, of the reference frame $\mathcal R_{\rm w}$ in which the turbulence is mostly magnetic. This frame moves at relativistic velocities towards the shock front at rest, decelerating gradually from the far to the near precursor~\cite{pap1,L1}. In a first part, we construct a fluid model to derive the deceleration law of the background plasma expected from the scattering of suprathermal particles off the microturbulence. This law leads to the relationship $\gamma_{\rm p}\,\sim\,\xi_{\rm b}^{-1/2}$ between the background plasma Lorentz factor $\gamma_{\rm p}$ and the normalized pressure of the beam $\xi_{\rm b}$; it is found to match nicely the spatial profiles observed in large-scale 2D3V particle-in-cell simulations. In a second part, we model the dynamics of the background plasma at the kinetic level, incorporating the inertial effects associated with the deceleration of $\mathcal R_{\rm w}$ into a Vlasov-Fokker-Planck equation for pitch-angle diffusion. We show how the effective gravity in $\mathcal R_{\rm w}$ drives the background plasma particles through friction on the microturbulence, leading to efficient plasma heating. Finally, we compare a Monte Carlo simulation of our model with dedicated PIC simulations and conclude that it can satisfactorily reproduce both the heating and the deceleration of the background plasma in the shock precursor, thereby providing a successful 1D description of the shock transition at the microscopic level.
\end{abstract}

\maketitle

\section{Introduction}\label{sec:introd}
The physics of collisionless shock waves is an important and long-standing theoretical question, which encompasses various fields of research, from fundamental plasma physics, to space plasma physics and laboratory astrophysics, see {\it e.g.}~\cite{Marcowith_2016} for a recent review. In high energy astrophysics, collisionless shock waves are -- nearly ubiquitously -- associated with the generation of nonthermal powerlaws of accelerated particles~\cite{1987PhR...154....1B}, whose secondary interactions with ambient fields provide the multi-messenger radiations that are being so actively studied~\cite{2011A&ARv..19...42B}. As the direct offsprings of the relativistic outflows of extremely powerful astrophysical sources, relativistic collisionless shock waves are associated with some of the most important outbursts of radiation seen in Nature, {\it e.g.}, gamma-ray bursts, pulsar wind nebulae, or blazar-type objects~\cite{2012SSRv..173..309B}.

In a fluid description, these shock waves are described as simple discontinuities, the quantities on both sides being related by the conservation laws of energy-momentum and matter fluxes, see~\cite{1976PhFl...19.1130B} for the relativistic regime. Particles that scatter off the magnetized flows on both sides of the shock front then gain energy through the so-called Fermi process, {\it e.g.}~\cite{1987PhR...154....1B,2015SSRv..191..519S} and references therein. However, it has long been recognized that one cannot apprehend this scenario in its inner workings without drawing a connection to the physics of the shock front and to the nature of the turbulence that is excited in the shock vicinity. In the rather spectacular case of the ultrarelativistic forward shock of a gamma-ray burst outflow, the magnetization of the unshocked plasma is so weak -- $\sigma\,\equiv\,(u_{\rm A}/c)^2\,\sim\,10^{-9}$ ($u_{\rm A}$ Alfv\'en four-velocity) -- that the accelerated particles actually scatter off a magnetized turbulence that they themselves excite through micro-instabilities in the shock precursor~\cite{1999ApJ...511..852G,1999ApJ...526..697M,2006ApJ...645L.129L,2008ApJ...682L...5S}.
The shock precursor is understood here as the region upstream of the shock where the beam of suprathermal/accelerated particles and the background plasma interpenetrate each other. 

In unmagnetized relativistic collisionless shocks, numerical particle-in-cell (PIC) simulations have demonstrated that the current filamentation instability (CFI)~\cite{Weibel_1959,Davidson_1972,2004PhRvE..70d6401B,2010PhRvE..81c6402B} is the dominant source of turbulence~\cite{2007ApJ...668..974K,2008ApJ...673L..39S,2008ApJ...682L...5S, 2009ApJ...695L.189M,2009ApJ...693L.127K, 2009ApJ...698L..10N,2013ApJ...771...54S}. The nature of the resulting turbulence, how it evolves in time and space, how it heats up the background plasma, how it scatters the suprathermal particles, and so forth, remain questions of importance which have been actively debated in the literature (see, {\it e.g.}~\cite{2004A&A...428..365W,2006ApJ...647.1250L,2007A&A...475....1A, 2007A&A...475...19A,2010MNRAS.402..321L,2011ApJ...736..157R, 2011MNRAS.418L..64L,2012EL.....9735002G,2012ApJ...744..182S, 2013MNRAS.430.1280P,Kumar_2015} or~\cite{2015SSRv..191..519S,2017SSRv..207..319P} for reviews in the relativistic limit).

This paper belongs to a series in which we discuss various physical aspects of unmagnetized, relativistic collisionless shocks (see~\cite{L1} for a summary of our model). Our theoretical analysis is supported by dedicated large-scale 2D3V PIC simulations of relativistic shocks in electron-positron plasmas. In Paper I~\cite{pap1}, we argue that there exists a frame $\mathcal R_{\rm w}$ in which the microturbulence that is excited in the precursor is mostly of magnetic nature, and we further show that this frame moves subrelativistically with respect to the background plasma. Building on these results, we address here the physics of deceleration and heating of the background plasma at a microscopic level.  In a subsequent Paper III~\cite{pap3} of this series, we discuss the physics of the suprathermal particles, in particular their scattering length and their distribution function in the shock precursor. Finally, in Paper IV~\cite{pap4}, we examine the microturbulence growth through current filamentation instabilities in the nonlinear regime, and provide comparisons to PIC simulations.

This paper is laid out as follows. In Sec.~\ref{sec:setup}, we set up the problem, explicit our approximations and define the main quantities of interest. In Sec.~\ref{sec:decel}, we discuss the deceleration of the background plasma imparted by the scattering of suprathermal particles. In Sec.~\ref{sec:heat}, we discuss the heating mechanism and compare its predictions to PIC simulations.  Finally, we summarize our results in Sec.~\ref{sec:conc}. Everywhere, we use Gaussian units with $k_{\rm B}\,=\,c\,=\,1$.

\section{Setup}
\label{sec:setup}
We discuss the dynamics of a background plasma as it penetrates the precursor of a relativistic collisionless shock in the lab frame $\mathcal R_{\rm s}$, which we define as that in which the shock front lies at rest at $x\,=\,0$. The background plasma moves along the $x-$axis with velocity $\beta_{\rm p}(x)\,<\,0$ (quantities indexed by $_{\rm p}$ refer hereafter to the background plasma). The asymptotic velocity of the background flow outside the precursor is written $\beta_\infty$ and its corresponding Lorentz factor $\gamma_\infty$ ($4-$velocity $u_\infty\,\equiv\,\gamma_\infty\beta_\infty$).  The precursor corresponds to the region $0\,<\,x\,<\,\ell_{\rm prec}$ that is permeated with suprathermal particles undergoing Fermi cycles around  the shock front. The length scale $\ell_{\rm prec}$ determines the maximum extent of this region (measured in the shock rest frame). Its value will be left unspecified for now, but for reference, in the time-dependent regime of a PIC simulation, the tip of the precursor moves at $\simeq c$ relative to the shock front, hence $\ell_{\rm prec}\,\simeq\, c t_{\rm max\vert d}$, where $t_{\rm max\vert d}$ is the integration time of the simulation. The bulk plasma velocity $\beta_{\rm p}$ and temperature $T_{\rm p}$ are both functions of the distance to the shock front. By virtue of the shock crossing conditions~\cite{1976PhFl...19.1130B}, $\beta_{\rm p}\,\rightarrow\,-1/3$ ($-1/2$ in 2D3V PIC simulations) and $T_{\rm p}\,\rightarrow\,\gamma_\infty m/(3\sqrt{2})$ as $x\,\rightarrow\,0$ [$\gamma_\infty m/(2\sqrt{3})$ in 2D3V PIC simulations], but $\beta_{\rm p}\,\rightarrow\,\beta_\infty$ and $T_{\rm p}\,\rightarrow\,0$ as $x\,\rightarrow\,\ell_{\rm prec}$. The main objective of the present work is to discuss the evolution of $\beta_{\rm p}$ and $T_{\rm p}$ as a function of $x$.

In the shock frame, the suprathermal particle distribution in momentum space, which we label with $_{\rm b}$ (b for ``beam'') is close to isotropic. Its characteristic proper temperature $T_{\rm b}$ is larger than the temperature of the shocked thermal plasma, as a result of the generation of an extended nonthermal tail. As in ~\cite{pap1,L1}, we write $T_{\rm b}\,=\,\kappa_{T_{\rm b}}\,\gamma_\infty m$, with $\kappa_{T_{\rm b}}$ of the order of a few to an order of magnitude. The pressure of this beam is conveniently scaled by $\xi_{\rm b}$, in terms of the momentum flux of the incoming plasma at infinity $F_\infty\,\equiv\,u_{\infty}^2w_\infty$:
\begin{equation}
\xi_{\rm b}\,\equiv\,\frac{p_{\rm b}}{F_\infty}\,.
\end{equation}
Here, $w_\infty\,\equiv\,n_{\infty}m$ (resp. $n_\infty$) corresponds to the proper enthalpy (resp. particle) density of the background plasma, which is assumed cold outside of the precursor. We assume that $\xi_{\rm b}$ depends on the distance to the shock front, with of course $\xi_{\rm b}\,\rightarrow\,0$ as $x\,\rightarrow\,\ell_{\rm prec}$ at the tip of the precursor.

We assume the shock to be unmagnetized, {\it i.e.} the background plasma does not carry any background magnetic field at $x\rightarrow+\infty$. Consequently, the CFI develops in the shock precursor through the interpenetration of the suprathermal and background plasma particle populations, thereby generating a small-scale electromagnetic turbulence. The transverse CFI mode produces plasma current filaments oriented along the shock normal, surrounded by a (toroidal) magnetic field $\boldsymbol{\delta}\mathbf{B}_{\boldsymbol{\perp}}$ and (radial) electric field $\boldsymbol{\delta}\mathbf{E}_{\boldsymbol{\perp}}$. As discussed in~\cite{pap1,L1}, this mode is essentially magnetic ({\it i.e.} mostly aperiodic, $\vert\omega\vert\,\ll\,k$) in some frame $\mathcal R_{\rm w}$, also denoted ``Weibel frame''. This frame is more specifically defined as that in which the transverse electrostatic component vanishes. In the frame $\mathcal R_{\rm s}$, this ``Weibel frame'' thus moves with velocity $\boldsymbol{\beta_{\rm w}}\,=\,\boldsymbol{\delta}\mathbf{E}_{\boldsymbol \perp}\times 
\boldsymbol{\delta}\mathbf{B}_{\boldsymbol \perp}/\delta B^2$.

We note that the ``Weibel frame'' may be ill-defined in some regions of the precursor, because the development of oblique modes may generate a configuration in which $\delta E_\perp\,\geq\,\delta B_\perp$. However, the transverse CFI mode is believed to dominate over most of the precursor, and especially close to the shock front, where deceleration and heating of the background plasma mainly take place, {\it e.g.}~\cite{2011MNRAS.418L..64L,2012ApJ...744..182S,2013MNRAS.430.1280P}. In~\cite{pap1,L1}, we have confirmed that this is indeed the case, {\it i.e.} that $\beta_{\rm w}$ is well defined within $\lesssim 300-1000\,c/\omega_{\rm p}$ of the precursor, using large-scale 2D3V PIC simulations. Yet the boost to $\mathcal R_{\rm w}$ cannot erase any longitudinal electric field component $\delta E_x$ that would originate from the growth of the CFI, or from a contribution of oblique modes. An important assumption that we make here is to neglect the influence of this longitudinal electric field. We will justify this assumption in more detail in Sec.~\ref{sec:dEx}. By contrast, the longitudinal electric field component is expected to play an important role in electron-ion shocks, because of the difference of inertia between negatively and positively charged species, regarding the deceleration as well as heating, {\it e.g.}~\cite{2008ApJ...673L..39S,Kumar_2015}.

In the case of symmetric counterstreaming configurations, the ``Weibel frame'' would coincide with the lab frame, $\beta_{\rm w}\,=\,0$. In the precursor of a shock, however, the suprathermal beam -- background plasma interaction is highly asymmetric, and therefore $\beta_{\rm w}\,\neq\,0$ in $\mathcal R_{\rm s}$~\cite{pap1,L1}. In particular, the suprathermal particles carry a high inertia in $\mathcal R_{\rm w}$ with a typical Lorentz factor $\sim\,\gamma_{\rm w}\gamma_{\infty}\,\gg\,1$, while the background plasma particles remain sub- or mildly relativistic over most of the precursor. Consequently, the suprathermal particles are not tied to the current filaments, but rather suffer small-angle deflections when crossing them. Their large scattering length sets the scale of variation of $\xi_{\rm b}$ in the $\mathcal R_{\rm s}$ frame. For these reasons, the microturbulence exerts a much larger influence on the background plasma than on the suprathermal particles.

Two other frames of interest are $\mathcal{R_{\rm p}}$, in which the plasma lies instantaneously at rest and $\mathcal R_{\rm d}$, which coincides with the rest frame of the downstream shocked plasma and with the reference frame of our numerical PIC simulations. Quantities indexed with a subscript $_{\rm\vert s,\,\vert w,\,\vert p,\,\vert d}$ are understood to be measured in the respective frame; frame-dependent quantities without a subscript are understood to be measured in the shock rest frame $\mathcal R_{\rm s}$.

\section{Deceleration of the background plasma}\label{sec:decel}
\subsection{General comments}\label{sec:genrem}
The present Section discusses the physics of deceleration of the background plasma inside the precursor in the framework of a fluid model, relying on the kinetic picture decribed in Paper~I~\cite{pap1}. In this picture, the ``Weibel frame'' $\mathcal R_{\rm w}$ has an everywhere nonvanishing relative velocity with respect to the background plasma as a consequence of the influence of suprathermal particles. At the same time, the background plasma keeps relaxing in $\mathcal R_{\rm w}$ through pitch-angle scattering in the magnetic turbulence. This permanent adjustment of the velocity of $\mathcal R_{\rm w}$ to the physical conditions at $x$ implies that $\mathcal R_{\rm w}$ keeps decelerating from large to small $x$, along the background plasma advection history in the shock frame.

In Paper~I~\cite{pap1}, the relative velocity $\beta_{\rm w\vert p}$ has been calculated in a homogeneous setting, for given conditions fixing the state of the suprathermal population and of the background plasma, using two main models: one that searches for a frame in which the dispersion relation of the CFI can be solved without an electrostatic component, and one that models the nonlinear phase of the CFI as a periodic collection of filaments in stationary pressure equilibrium. We have then shown that these models provide a satisfactory fit to the spatial profile of $\beta_{\rm w\vert p}$ extracted from PIC simulations, if the model predictions are calculated at point $x$ using the physical conditions of the populations at this point. 

In principle, to obtain a self-consistent fully kinetic model of this deceleration process, one would need to extend the calculations of Paper I to a non-stationary state, and to include in this time-dependent framework the physics of relaxation of the background plasma; this  formidable task lies well beyond the scope of any current study, however. To overcome this difficulty, we construct here a fluid analog of the above microscopic picture. The influence of suprathermal particles is, in particular, characterized by their kinetic pressure. We then solve the equations of conservation of energy-momentum to obtain the deceleration law as a function of the profile $\xi_{\rm b}(x)$. We also evaluate the influence of the microturbulence on the background plasma and conclude that it can be safely neglected here: at the fluid level, this means that the turbulence is effectively tied to the background plasma, in good accord with $\vert\beta_{\rm w\vert p}\vert\,\ll\,1$. Note that the buildup of the turbulence in the precursor through the CFI also exerts work on the streaming background plasma, but arguments developed in Paper~IV~\cite{pap4} indicate that this ponderomotive force is negligible. 

\subsection{The fluid law of deceleration}\label{sec:declaw}
In order to derive an estimate for the deceleration law of the $\mathcal R_{\rm w}$ frame, we use a fluid description of the conservation of energy and momentum in the plasma $+$ electromagnetic turbulence $+$ beam system. We describe the background plasma as a perfect fluid, with velocity $\beta_{\rm p}$, enthalpy $w_{\rm p}$, pressure $p_{\rm p}$ and equation of state $w_{\rm p}\,=\,n_{\rm p}m+\alpha_{\rm p}p_{\rm p}$, with the short-hand notation $\alpha_{\rm p}\,\equiv\,\hat\Gamma_{\rm p}/\left(\hat\Gamma_{\rm p}-1\right)$ (and similarly for the beam). Current conservation implies $\gamma_{\rm p}\beta_{\rm p}n_{\rm p}\,=\,\gamma_\infty\beta_\infty n_\infty$.

Since both the beam and turbulence energy and momentum densities vanish outside the precursor, one can write the integrated version of the equations of energy-momentum conservation between as $+\infty$ and a point $x$ as
\begin{eqnarray}
\gamma_\infty^2\beta_\infty w_\infty&\,=\,&
\gamma_{\rm p}\gamma_\infty\beta_\infty w_\infty + \gamma_{\rm p}^2\beta_{\rm p}\alpha_{\rm p}p_{\rm p}+ {T_{\rm b}}^{tx} + {T_B}^{tx}\,,\nonumber\\
\gamma_\infty^2\beta_\infty^2 w_\infty &\,=\,& \gamma_{\rm p}\beta_{\rm p}\gamma_\infty\beta_\infty w_\infty + \left(\gamma_{\rm p}^2\beta_{\rm p}^2\alpha_{\rm p}+1\right)p_{\rm p}+ {T_{\rm b}}^{xx}\nonumber\\
&&\quad\quad  + {T_B}^{xx}\,.
\label{eq:eqfl}
\end{eqnarray}
We simplify the notations by defining ${T_{\rm b}}^{tx}\,\equiv\,\Psi_{\rm b}F_\infty/\beta_\infty$,  ${T_{\rm b}}^{xx}\,\equiv\,\Phi_{\rm b}F_\infty$,  ${T_B}^{tx}\,\equiv\,\Psi_B F_\infty/\beta_\infty$ and  ${T_B}^{xx}\,\equiv\,\Phi_B F_\infty$. Finally, we also define $\Psi\,=\,\Psi_{\rm b}+\Psi_B$, $\Phi\,=\,\Phi_{\rm b}+\Phi_B$ so that the above system can be rewritten as  
\begin{eqnarray}
\frac{\gamma_{\rm p}}{\gamma_\infty}F_\infty + \gamma_{\rm p}^2\beta_{\rm p}\beta_\infty\alpha_{\rm p}p_{\rm p} &\,=\,&F_\infty\left(1-\Psi\right)\,,\nonumber\\
\frac{\gamma_{\rm p}\beta_{\rm p}}{\gamma_\infty\beta_\infty}F_\infty + \left(\gamma_{\rm p}^2\beta_{\rm p}^2\alpha_{\rm p}+1\right)p_{\rm p}&\,=\,&F_\infty\left(1-\Phi\right)\,.
\label{eq:eqfl2}
\end{eqnarray}
In this form, $-\Psi$ represents the net fraction of initial energy density picked up by the background plasma on its way, relative to the incoming momentum flux at infinity, while $-\Phi$ similarly represents the relative fraction of picked up momentum density. It proves instructive to explicit these terms, notably to show that $\vert\Psi\vert\,\ll\,1$ and $\vert\Phi\vert\,\ll\,1$:
\begin{eqnarray}
\Psi_{\rm b}&\,=\,&\gamma_{\rm b}^2\beta_{\rm b}\beta_\infty\alpha_{\rm b}\xi_{\rm b}\,,\nonumber\\
\Phi_{\rm b}&\,=\,&\left(\gamma_{\rm b}^2\beta_{\rm b}^2\alpha_{\rm b}+1\right)\xi_{\rm b}\,,\nonumber\\
\Psi_B&\,=\,& \frac{\gamma_{\rm w}^2\beta_{\rm w}\beta_\infty}{\gamma_\infty^2\beta_\infty^2}\,2\,\epsilon_B\,,\nonumber\\
\Phi_B&\,=\,& \frac{\gamma_{\rm w}^2\beta_{\rm w}^2}{\gamma_\infty^2\beta_\infty^2}\left(2 +\frac{1}{\gamma_{\rm w}^2\beta_{\rm w}^2}\right)\epsilon_B\,,
\label{eq:eqfl3}
\end{eqnarray}
with $\xi_{\rm b}\,\ll\,1$ and $\epsilon_B\,\ll\,1$. Above, we have assumed a perfect fluid form for the beam energy-momentum (and neglected rest-mass energy in front of the pressure of suprathermal particles). Moreover, the energy-momentum tensor of the microturbulence takes on a relativistic MHD form, since the electric field vanishes in the $\mathcal R_{\rm w}$ frame by definition (and the above assumes a magnetic field transverse to the flow). Finally, $\beta_{\rm w}\,\simeq\,\beta_{\rm p}$ in the relativistic regime, since
\begin{equation}
\beta_{\rm w}\,\simeq\,\beta_{\rm p}\left(1+\frac{\beta_{\rm p\vert w}}{\gamma_{\rm p}^2}\right)\,,
\label{eq:eqfl4}
\end{equation}
to first order in $1/\gamma_{\rm p}^2$ if $\vert\beta_{\rm p\vert w}\vert\,\ll\,1$.

The above equations imply
\begin{equation}
p_{\rm p}\,=\,F_\infty\left(\frac{\beta_{\rm p}}{\beta_\infty}\Psi-\Phi + 1 - \frac{\beta_{\rm p}}{\beta_\infty}\right)\,.
\label{eq:eqfl5}
\end{equation}
Since $\beta_{\rm p}/\beta_\infty\,\simeq\,1 - 1/(2\gamma_{\rm p}^2)+1/(2\gamma_\infty^2)$ as long as $\gamma_{\rm p}\,\gg\,1$, Eq.~(\ref{eq:eqfl5}) implies that $p_{\rm p}/w_\infty$ becomes of order unity or larger as soon as $\vert\Phi-\Psi\vert\,\gtrsim\,1/\gamma_\infty^2$. Once this inequality is satisfied, the background plasma effectively slows down in the precursor.

This relationship is by itself not surprising: when $\Phi\,\rightarrow\,0$, meaning that the plasma picks up matter at rest in the laboratory frame, it is well-known that it suffices to add in a fraction $\sim\,1/\gamma_\infty^2$ of the incoming energy to slow down the plasma, because once picked up by the flow, the supplementary mass-energy is increased by a factor $\gamma_\infty^2$~\cite{1976PhFl...19.1130B} (see also \cite{2016MNRAS.460.2036D} for a recent discussion on similar issues). However, if $\vert\Psi-\Phi\vert\,\lesssim\,1/\gamma_\infty^2$, the loading amounts to adding matter moving at about the same velocity as the background plasma, in which case there is no increase by $\gamma_\infty^2$.

This result does confirm that the microturbulence exerts a negligible influence on the background plasma, as indeed
\begin{equation}
\Phi_B-\Psi_B\,\simeq\,\frac{\epsilon_B}{\gamma_{\rm p}^2}\,,
\end{equation}
according to Eqs.~(\ref{eq:eqfl3}) and (\ref{eq:eqfl4}). Let us stress that this result rests on the observation that $\vert\beta_{\rm p\vert w}\vert\,\ll\,1$, which implies that the turbulence is essentially carried by the background plasma. In the following, we thus neglect the contribution of the microturbulence and retain only that of the beam. In this limit, the fluid model becomes the relativistic generalization of ``cosmic-ray modified'' shocks~\cite{1981ApJ...248..344D,1987PhR...154....1B}, which have been observed in numerical nonlinear Monte Carlo simulations of shock acceleration in the relativistic limit~\cite{2002APh....18..213E,*2004APh....22..323E,*2013ApJ...776...46E,*2015MNRAS.452..431W,*2016MNRAS.456.3090E}.

Since $\vert\Phi_{\rm b}-\Psi_{\rm b}\vert\,=\,\mathcal O(\xi_{\rm b})$, and $1-\beta_{\rm p}/\beta_\infty\,=\,\mathcal O\left(\gamma_{\rm p}^{-2}\right)$,
Eq.~(\ref{eq:eqfl5}) implies $\gamma_{\rm p}\,\simeq\,\gamma_\infty$ if $\gamma_\infty^2\xi_{\rm b}\,\ll\,1$. In the opposite limit, deceleration takes place; in particular, in the limit $1\,\ll\,\gamma_{\rm p}\,\ll\,\gamma_\infty$, using Eq.~(\ref{eq:eqfl5}) in Eq.~(\ref{eq:eqfl2}) and neglecting terms of order $\gamma_{\rm p}/\gamma_\infty$, $1/\gamma_{\rm p}^2$ or $\Psi$ in front of unity, one obtains 
\begin{equation}
\alpha_{\rm p}\gamma_{\rm p}^2\left(\frac{1}{2\gamma_{\rm p}^2}+\Psi_{\rm b}-\Phi_{\rm b}\right)\,\simeq\,1 \quad\quad\left(\gamma_\infty^2\xi_{\rm b}\,\gg\,1\right)\,,
\label{eq:eqfl6a}
\end{equation}
which provides the deceleration law
\begin{equation}
\gamma_{\rm p}\,\simeq\,\left[\frac{\alpha_{\rm p}-2}{2\alpha_{\rm p}\left(\Phi_{\rm b}-\Psi_{\rm b}\right)}\right]^{1/2} \quad\quad\left(\gamma_\infty^2\xi_{\rm b}\,\gg\,1\right)\,,
\label{eq:eqfl6}
\end{equation}
or, in terms of $\xi_{\rm b}$,
\begin{eqnarray}
\gamma_{\rm p}&\,\simeq\,&\gamma_\infty\quad\quad\left(\gamma_\infty^2\xi_{\rm b}\,\ll\,1\right)\,,\nonumber\\
\gamma_{\rm p}&\,\simeq\,&\left\{\frac{\alpha_{\rm p}-2}{2\alpha_{\rm p}\left[1+\gamma_{\rm b}^2\beta_{\rm b}(1+\beta_{\rm b})\alpha_{\rm b}\right]}\right\}^{1/2}\xi_{\rm b}^{-1/2}\nonumber\\
&&\quad\quad\left(\gamma_\infty^2\xi_{\rm b}\,\gg\,1\right)\,.
\label{eq:eqfl7}
\end{eqnarray}
In terms of $\beta_{\rm p}$, this law can be rewritten
\begin{eqnarray}
\beta_{\rm p}&\,\simeq\,&\beta_\infty\quad\quad\left(\gamma_\infty^2\xi_{\rm b}\,\ll\,1\right)\,,\nonumber\\
\beta_{\rm p}&\,\simeq\,&\beta_\infty\left\{1 - \frac{\alpha_{\rm p}\left[1+\gamma_{\rm b}^2\beta_{\rm b}(1+\beta_{\rm b})\alpha_{\rm b}\right]}{\alpha_{\rm p}-2}\xi_{\rm b}\right\}\nonumber\\
&&\quad\quad\left(\gamma_\infty^2\xi_{\rm b}\,\gg\,1\right)\,.
\label{eq:eqfl8}
\end{eqnarray}
 
\begin{figure}
\includegraphics[width=0.43\textwidth]{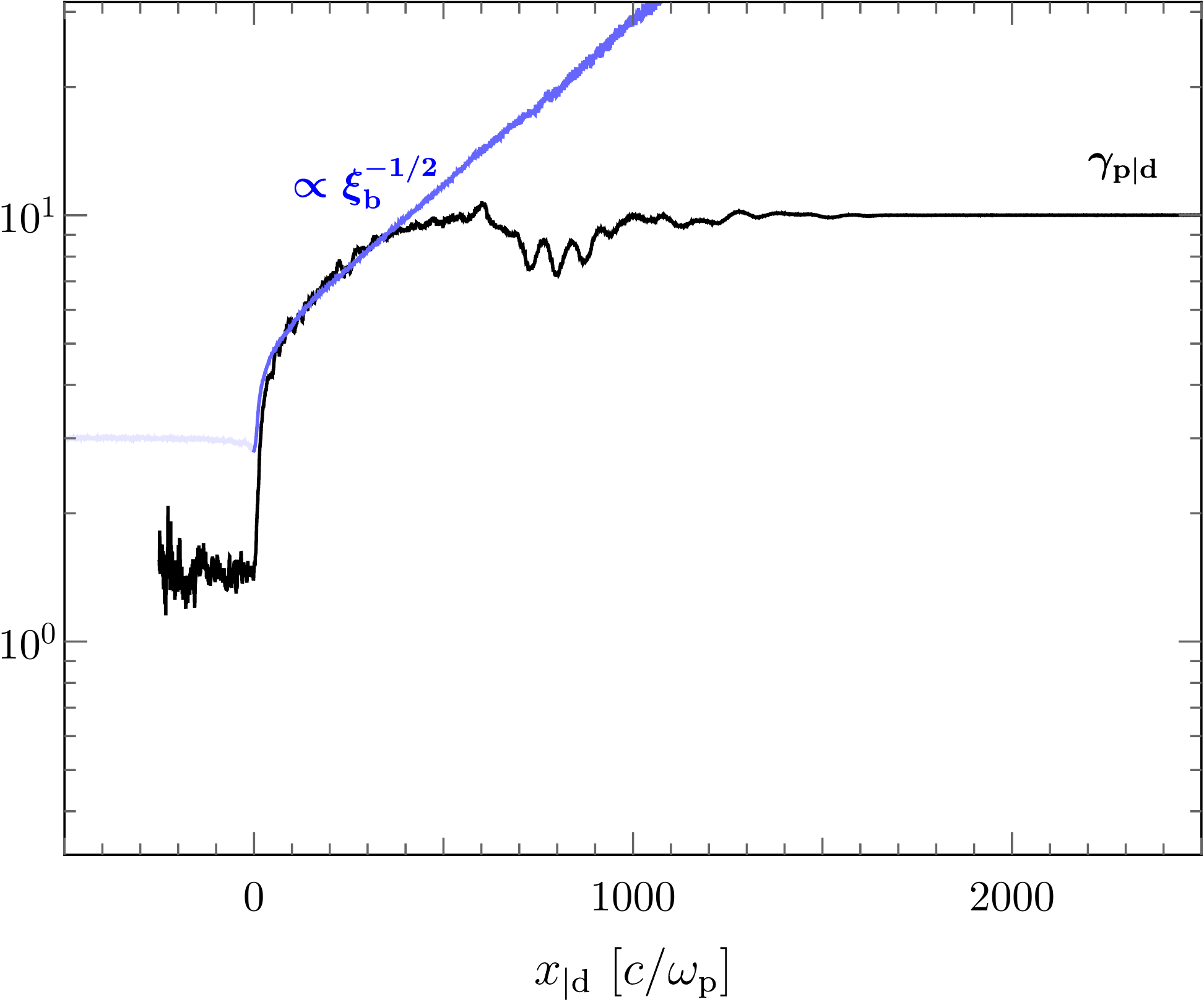}\smallskip\\
\includegraphics[width=0.43\textwidth]{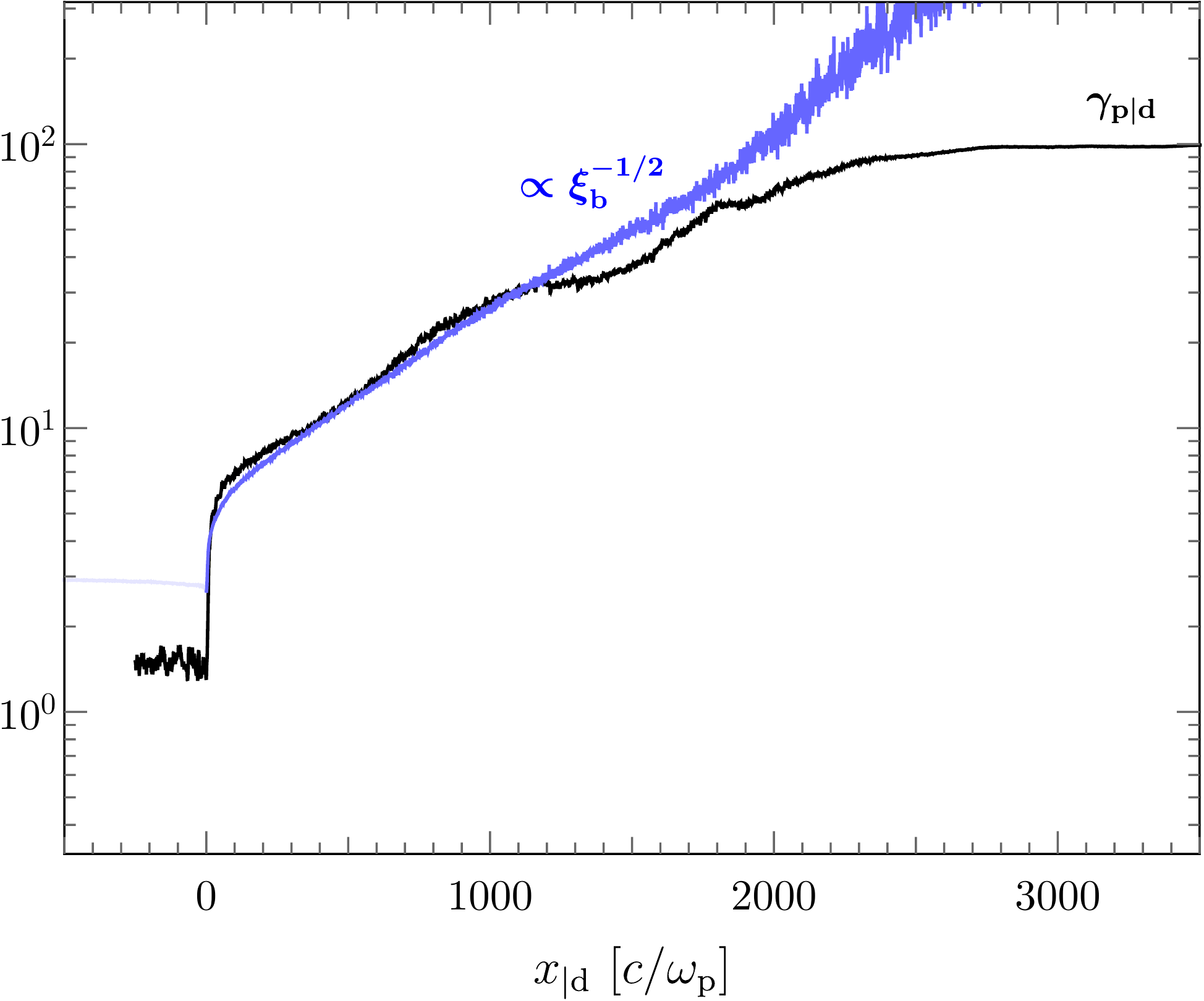}
  \caption{In black, the Lorentz factor of the background plasma, $\gamma_{\rm p\vert d}$, {\it vs} distance to the shock, extracted from PIC simulations of a pair shock of Lorentz factor $\gamma_{\infty\vert\rm d}\,=\,10$ (top panel) and $\gamma_{\infty\vert\rm d}\,=\,100$ (bottom panel). The law $\gamma_{\rm p\vert d}\,=\,1.2\,\xi_{\rm b}^{-1/2}$ is overplotted in blue, using as input the profile $\xi_{\rm b}(x)$ extracted from the same PIC simulations and the same {\it ad hoc} numerical prefactor $1.2$ in both panels. The law $\gamma_{\rm p\vert d}\,\sim\,\xi_{\rm b}^{-1/2}$ is that predicted by Eq.~(\ref{eq:eqfl7}) in the relativistic decelerating regime ($1\,\ll\,\gamma_{\rm p}\,\ll\,\gamma_\infty$). The numerical data is light colored in regions where it cannot be measured accurately.}
  \label{fig:xivsg_new} 
\end{figure}

In Fig.~\ref{fig:xivsg_new}, we compare the law $\gamma_{\rm p\vert d}\,\sim\,\xi_{\rm b}^{-1/2}$ in the deceleration regime ($\xi_{\rm b}\,\gtrsim\,1/\gamma_\infty^2$ or $\gamma_{\rm p}\,\lesssim\,\gamma_\infty$) with the profile of $\gamma_{\rm p\vert d}$ measured in two PIC simulations with Lorentz factors $\gamma_\infty\,=\,17$ and $\gamma_\infty\,=\,173$, which  correspond, respectively, to Lorentz factors $\gamma_{\infty\vert\rm d}\,=\,10$ and $100$ in the simulation frame $\mathcal R_{\rm d}$. These large-scale 2D3V simulations of unmagnetized, relativistic collisionless pair plasma shocks have been carrried out using the parallelized finite difference time-domain PIC code \textsc{calder}~\cite{Lefebvre_2003}.  The plasma is injected at the right-hand boundary of the simulation box with a proper temperature $T_\infty/m = 10^{-2}$ and reflects on a conducting wall to form the shock~\cite{2008ApJ...682L...5S}. Each simulation initially contains 10 macro-particles per cell and per species; the cell size is $\Delta x \,=\,\Delta y\,=\, 0.1\,c/\omega_{\rm p}$. The \v Cerenkov instability is mitigated using the Godfrey-Vay filtering algorithm~\cite{Godfrey_2014} and the Cole-Karkkainen finite difference field solver~\cite{Cole_1997a, Cole_2002, Karkkainen_2006}, which allows a large time-step to be used, $\Delta t = 0.99 \Delta x /c$. Dedicated particle diagnostics have been implemented to distinguish the beam particles from the background plasma: by our definition, background plasma particles correspond to those particles whose velocity along $\bm{x}$ is negative and has never changed sign; by contrast, beam particles are defined as those particles whose velocity along $\bm{x}$ is positive. We note here that the $4-$velocity and the temperature of the background plasma cannot be extracted accurately from the PIC simulation beyond the shock transition, since its particles have by then been isotropized, and hence they can no longer be distinguished from the suprathermal particles.

 In Fig.~\ref{fig:xivsg_new}, the profile of $\xi_{\rm b}^{-1/2}$ has been multiplied by an {\it ad hoc} factor of the order of unity ($1.5$) to match the profile of $\gamma_{\rm p}$ in the corresponding region. Equation~(\ref{eq:eqfl7}) indeed suggests a prefactor of the order of unity, but which is difficult to estimate in our model because of an uncertainty related to the value of $\beta_{\rm b}$. In Paper~III, we discuss this latter quantity and show that $\beta_{\rm b}$ is small in magnitude, possibly negative, in the shock front frame, but it is not accurately defined, because the theoretical solution for $\xi_{\rm b}$ in that study neglects the deceleration of the background plasma as well as a possible evolution of the scattering length with $x$.

\begin{figure}
\includegraphics[width=0.45\textwidth]{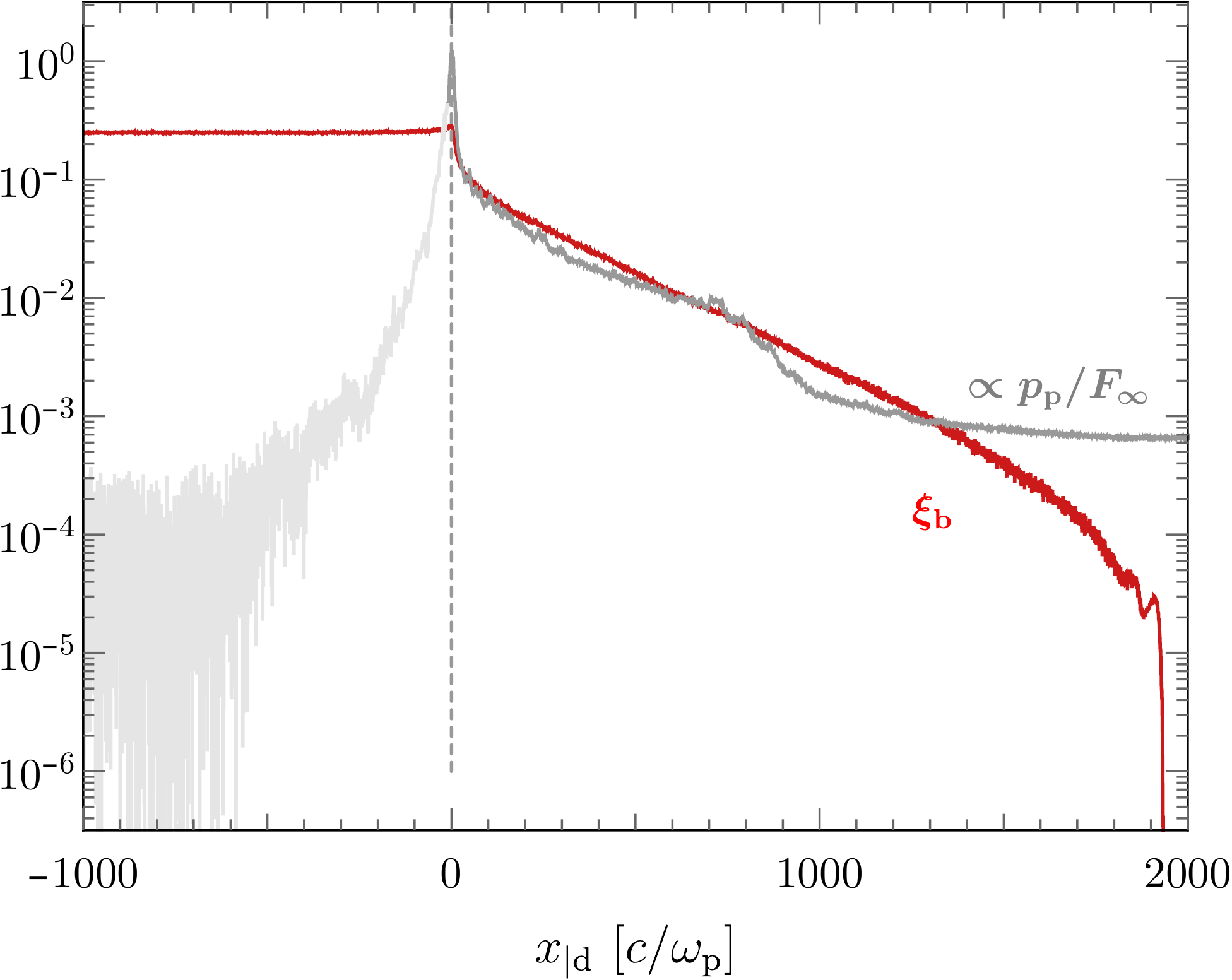}\smallskip\\
\includegraphics[width=0.45\textwidth]{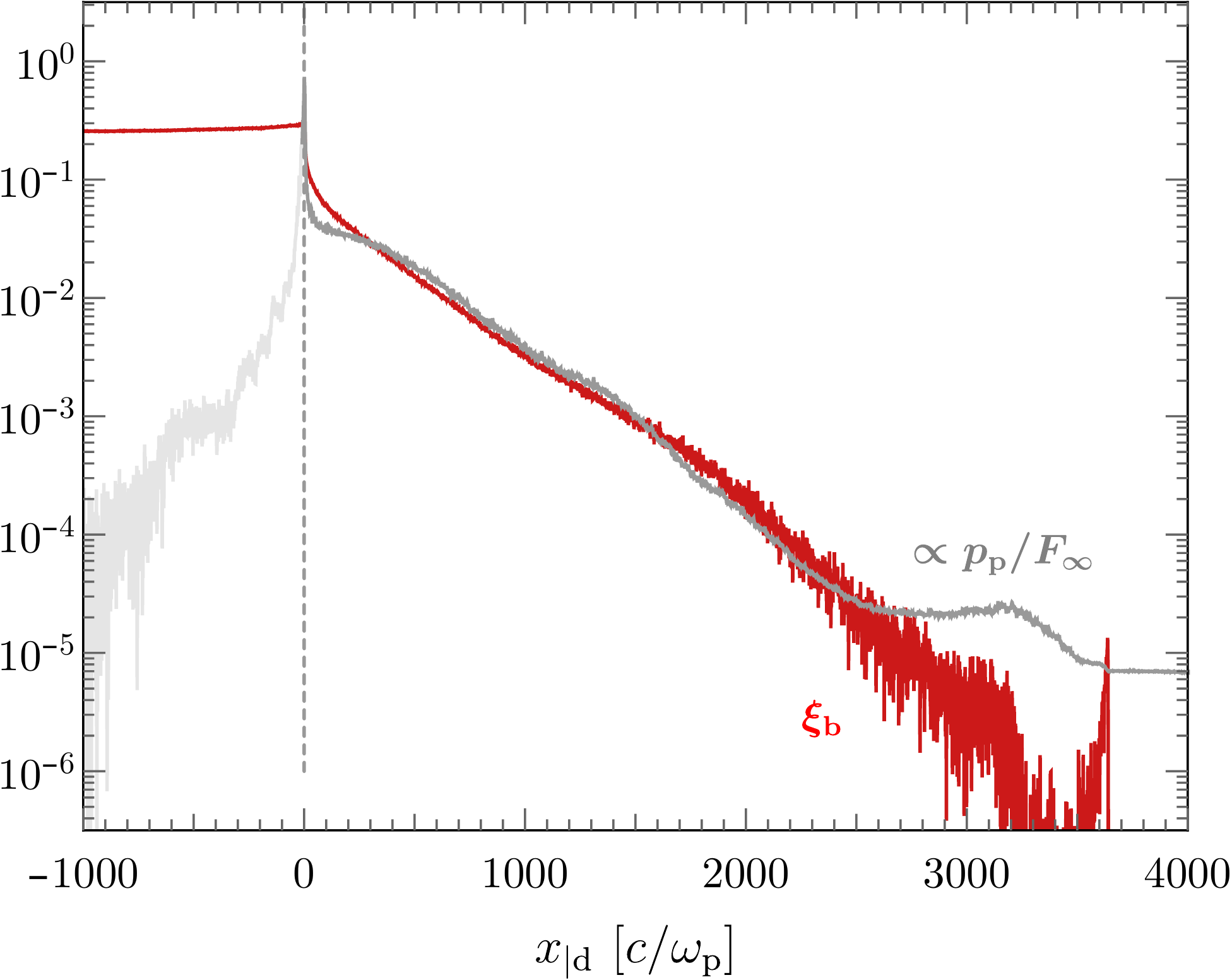}
\caption{In red, the normalized pressure of suprathermal particles $\xi_{\rm b}$ vs the distance to the shock, as extracted from PIC simulations of a pair shock of Lorentz factor $\gamma_{\infty\vert\rm d}\,=\,10$ (top panel) and $\gamma_{\infty\vert\rm d}\,=\,100$ (bottom panel). The law $(0.15)^{-1}\,p_{\rm p}/F_\infty$ is overplotted in gray, using the same {\it ad hoc} numerical prefactor $(0.15)^{-1}$ in both panels. This scaling $p_{\rm p}\,\propto\,F_\infty\xi_{\rm b}$ is that predicted by Eq.~(\ref{eq:eqpp}) in the relativistic decelerating regime ($1\,\ll\,\gamma_{\rm p}\,\ll\,\gamma_\infty$). As before, data is light colored in regions where it cannot be measured accurately.
  \label{fig:xibvsp_new} }
\end{figure}

We note that the numerical prefactor takes on a similar value in both simulations, despite different $\gamma_\infty$ values. Furthermore, the profiles of $\gamma_{\rm p \vert d}$ and $\xi_{\rm b}^{-1/2}$ from both simulations appear to be closely similar over the same deceleration zones. This suggests that, in the deceleration region, the precursor obeys a universal law $\gamma_{\rm p}/\gamma_\infty$ vs $x$, or $\xi_{\rm b}(x)$ vs $x$. 

Using Eq.~(\ref{eq:eqfl7}) in the solution obtained for the pressure, Eq.~(\ref{eq:eqfl5}), we derive the additional scaling law 
\begin{equation}
p_{\rm p}\,\simeq\,2F_\infty\frac{\Phi-\Psi}{\alpha_{\rm p}-2}\,\propto\,F_\infty\,\xi_{\rm b}\quad \left(\gamma_\infty^2\xi_{\rm b}\,\gg\,1\right)\,.
\label{eq:eqpp}
\end{equation}
In Fig.~\ref{fig:xibvsp_new}, we compare this law to the numerical data obtained in our PIC simulations, and find that $p_{\rm p}\simeq 0.15 F_\infty\xi_{\rm b}$ provides a satisfactory agreement. We note that this prefactor is not independent from that describing the relationship between $\gamma_{\rm p}$ and $\xi_{\rm b}$. Specifically, writing $p_{\rm p}\,=\,a_{\rm p}F_\infty\xi_{\rm b}$ and $\gamma_{\rm p}\,=\,a_{\gamma}\xi_{\rm b}^{-1/2}$, one finds from Eqs.~(\ref{eq:eqfl7}) and (\ref{eq:eqfl5}): $a_{\rm p}=\left(\alpha_{\rm p}a_{\gamma}^2\right)^{-1}$. Hence, for $a_\gamma\simeq1.5$ and $\alpha_{\rm p}\simeq3$ (hot gas in 2D), there results $a_{\rm p}\simeq0.15$.

\subsection{The shock transition}
The above model nicely explains the value $\xi_{\rm b}\,\sim\,0.1$ of the fraction of energy injected into the suprathermal particle power-law tail, as repeatedly observed in PIC simulations, regardless of the Lorentz factor $\gamma_\infty$, {\it e.g.}~\cite{2013ApJ...771...54S}. Where $\xi_{\rm b}\,\simeq\,0.1$, indeed, the law of deceleration implies that $\gamma_{\rm p}$ becomes of the order of unity, {\it i.e.} the flow becomes subrelativistic and the shock transition forms over a length scale of $\sim100\,\omega_{\rm p}^{-1}$. Hence, that $\xi_{\rm b}\,\sim\,0.1$ in the immediate precursor is a natural prediction of our model.

This deceleration law plotted in Fig.~\ref{fig:xivsg_new} also reveals the existence of a sub-shock, which is a generic prediction of cosmic-ray modified shocks\footnote{We thank A. Levinson for pointing this out to us.}: the Lorentz factor of the background plasma is seen to slowly decrease in the precursor over some $\sim10^3\,\omega_{\rm p}^{-1}$ down to a value $\gamma_{\rm sub}\,\sim\,5$, at which point the shock transition occurs abruptly. This sub-shock arises here as a result of the transition from ultrarelativistic to mildly relativistic flow velocities. Far in the shock precursor, the length scales characteristic of the background plasma dynamics are typically dilated by a factor $\gamma_{\rm p}$ when expressed into the shock frame, so that the transition from ultrarelativistic to subrelativistic velocities implies a rapid evolution of the various physical quantities. For instance, the typical relaxation length scale of background plasma particles in the microturbulence is of the order of $\gamma_{\rm w}/\nu_{\vert\rm w}\,\gg\,\nu_{\vert\rm w}^{-1}$ in the shock frame, as discussed in the forthcoming Sec.~\ref{sec:heat}. Therefore, it drops to $\nu_{\vert\rm w}^{-1}$ once $\xi_{\rm b}\,\gtrsim\,0.1$. For the typical value $\nu_{\vert \rm w} \,\sim\,0.01\omega_{\rm p}$ inferred in Sec. IV, the relaxation scale is found to be of $\sim 100\,\c/\omega_{\rm p}$ in the shock transition. Similarly, the scattering length of beam particles of Lorentz factor $\gamma$ is of the order of $\gamma_{\rm w}\epsilon_B^{-1}(\gamma/\gamma_\infty)^2\,\omega_{\rm p}^{-1}$~\cite{pap3}, which becomes shorter by a factor $\gamma_{\rm w}$ once the Weibel frame has slowed to subrelativistic velocities.

  The above model neglects the difference between $\beta_{\rm w}$ and $\beta_{\rm p}$. This is well justified in the relativistic regime $\gamma_{\rm p}\,\gg\,1$, see Eq.~(\ref{eq:eqfl4}), but this approximation fails in the subrelativistic regime $\beta_{\rm p}\,\lesssim\,1$. In these last hundreds of skin depths, as discussed in Sec.~\ref{sec:heat} and in Paper~I, the plasma decouples from the microturbulence, before eventually relaxing once the $\mathcal R_{\rm w}$ has decelerated to a constant velocity $\beta_{\rm w}\,\sim\,-1/2$ (in 2D). The resulting isotropization of the background plasma injects into the upstream a fraction of particles, which then populate the suprathermal population. From their point of view, the deceleration of the background plasma builds up a scattering barrier over a few tens of $c/\omega_{\rm p}$: this scattering barrier results from the reduction of the scattering length scale due to the reduction in $\gamma_{\rm w}$ and the increase in $\epsilon_B$, relatively to values seen further in the precursor. Hence, those particles that are energetic enough to cross the barrier are free to stream into the precursor over long distances, and to populate the suprathermal particle tail.

\begin{figure}
\includegraphics[width=0.45\textwidth]{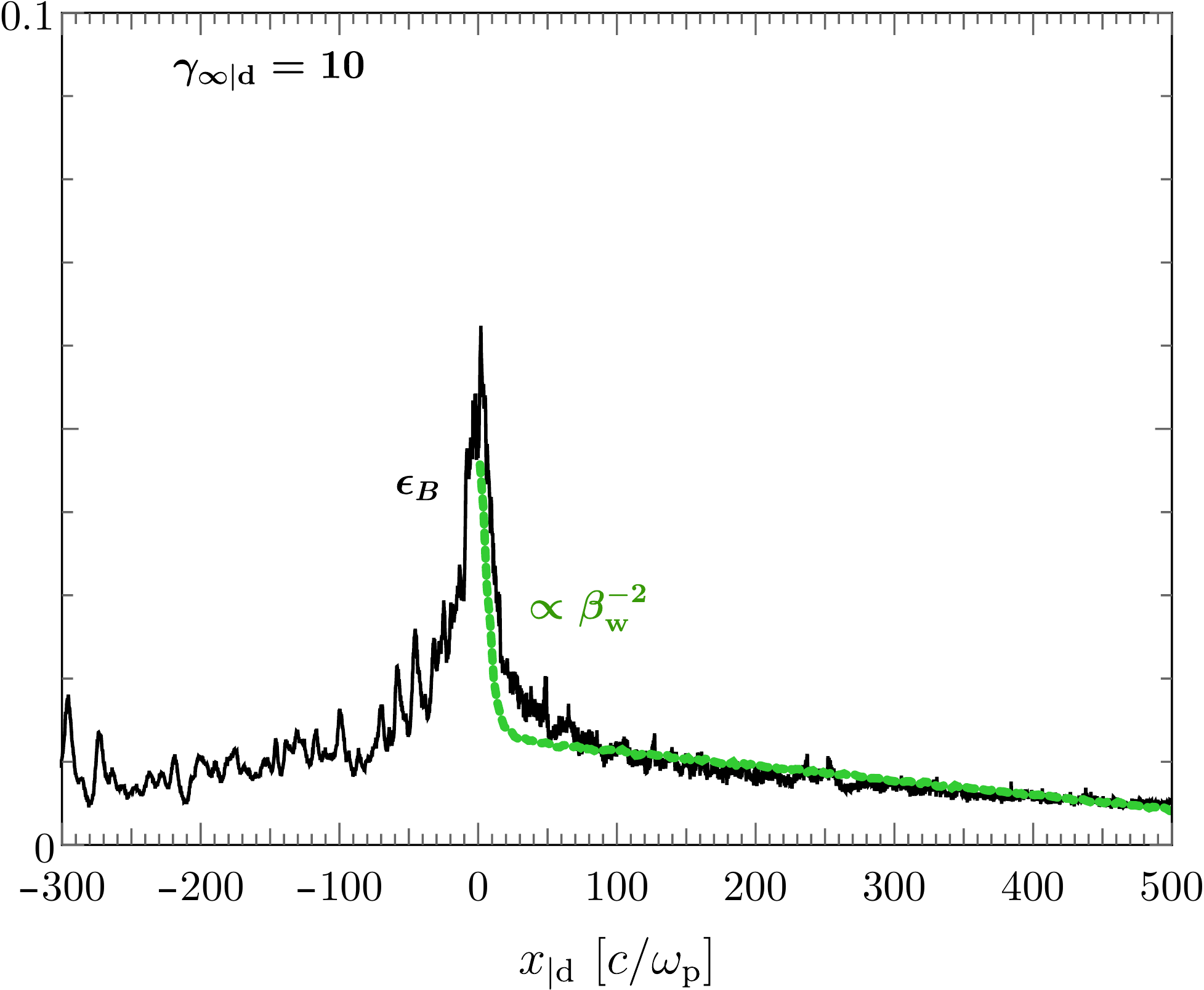}\smallskip\\
\includegraphics[width=0.45\textwidth]{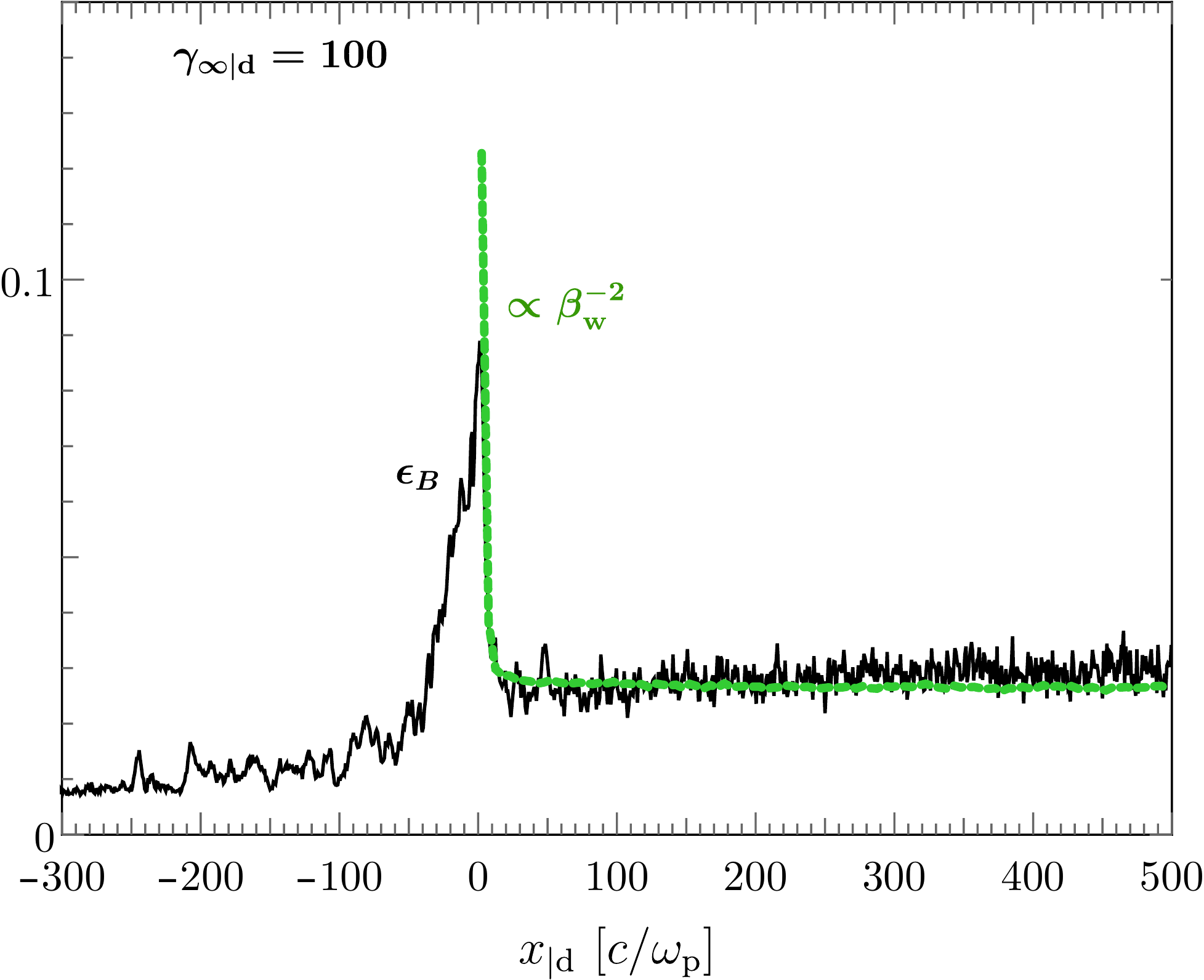}
  \caption{Evolution of the microturbulence strength parameter $\epsilon_B$ in the near precursor, as a function of distance to the shock, extracted from PIC simulations of a pair shock of Lorentz factor $\gamma_{\infty\vert\rm d}\,=\,10$ (top panel) and $\gamma_{\infty\vert\rm d}\,=\,100$ (bottom panel), in black. The MHD-like compression law $\epsilon_B\,\propto\,1/\beta_{\rm w}^2$, with $\beta_{\rm w}$ extracted from the same PIC simulations, is overplotted in dashed green, for comparison. The proportionality factor is chosen to match the $\epsilon_B(x_{\vert\rm d})$ curve in the precursor at distances $x_{\vert\rm d}\,\gtrsim100\,c/\omega_{\rm p}$  in order to remove its possible spatial dependence. This choice corresponds to $\epsilon_B\,=\,\left(1.2\times 10^{-2}-1.6\times 10^{-5}x_{\rm \vert d}\right)\,\beta_{\rm w}^{-2}$ in the top panel and $\epsilon_B\,=\,0.026\,\beta_{\rm w}^{-2}$ in the bottom panel.
  \label{fig:epsBvsx} }
\end{figure}

  Another prediction of the above model is that the strength of the microturbulence should increase as a result of the compression of magnetic field lines, modulo the growth factor imposed by the CFI, once the velocity of the $\mathcal R_{\rm w}$ frame becomes subrelativistic. This effect is best seen using Faraday's law in $\mathcal R_{\rm s}$, in steady state: $\boldsymbol{\nabla}\times\boldsymbol{\delta}\mathbf{E}=0$, which implies $\partial_x\delta E_\perp=0$ in the absence of a longitudinal $\delta E_x$ component. Note that, assuming $\delta E_x=0$ implicitly means neglecting the possible growth of the instability in the vicinity of the shock, which is indeed a reasonable assumption (see below). Since $\delta E_\perp=\beta_{\rm w}\delta B_\perp$, the above law implies $\epsilon_B\,\sim\,\epsilon_{B}^{\rm (far)}/\beta_{\rm w}^2$ in $\mathcal R_{\rm s}$, with $\epsilon_B^{\rm (far)}$ representing a typical value of $\epsilon_B$ well into the precursor.  We compare this prediction with PIC simulations of a pair shock of Lorentz factor $\gamma_{\infty\vert\rm d}\,=\,10$ and $\gamma_{\infty\vert\rm d}\,=\,100$ in Fig.~\ref{fig:epsBvsx}. Specifically, we overplot on the $\epsilon_B$ of both PIC simulations an {\it ad hoc} prefactor times $1/\beta_{\rm w}^2$, where $\beta_{\rm w}$ is inferred through the ratio $\langle E_y^2\rangle^{1/2}/\langle B_z^2\rangle^{1/2}$ from these simulations, see~\cite{pap1,L1}. In the $\gamma_{\infty\vert\rm d}\,=\,10$ case, the profile of $\epsilon_B$ in the upstream suggests that the CFI is still growing close to the shock, albeit weakly so. Notwithstanding this growth, the predicted profile due to magnetic compression provides a reasonable match to the observed peak of $\epsilon_B$. 

In the standard picture in which the CFI excites the growth of a magnetic barrier that peaks at the shock, one nagging question is: given that the growth takes from the tip of the precursor (at $x\,=\,\ell_{\rm prec}$) until the shock front (at $x\,=\,0$), why would the CFI yield the ``right'' value of $\epsilon_B$ at the location predicted by a fluid description of the shock front, to induce there the shock transition? This question has a simple and self-consistent answer in the present model: the peak in $\epsilon_B$ is not associated with an explosion of the instability but, rather, with the compression of the flow, which occurs where $\beta_{\rm w}$ turns subrelativistic, equivalently where the profile $\xi_{\rm b}(x)$ finds its maximum, {\it i.e.}, at the shock transition. 

Downstream of this peak, one does not expect the compression law to hold anymore, since it is known that the magnetic field relaxes through dissipation on short length scales in the downstream. Comparison of the $\beta_{\rm w}^{-2}$ and $\epsilon_B$ profiles suggests that dissipation washes out the magnetic energy by a factor of $3-4$. \\

\section{noninertial heating of the background plasma}\label{sec:heat}
In the present section, we describe the heating physics of the background
plasma through its interaction with a microturbulence that can be described as magnetostatic in the frame $\mathcal R_{\rm w}$, but which moves with a non-uniform bulk velocity $\beta_{\rm w}(x)$ in the lab-frame $\mathcal R_{\rm s}$. 

\subsection{Main equations}
Given that pitch-angle scattering takes place in the $\mathcal R_{\rm w}$ frame, while we seek to construct a stationary model in the $\mathcal R_{\rm s}$ frame, it proves convenient to write down the equation for the distribution function of the background plasma in a mixed coordinate system, with space variables defined in the $\mathcal R_{\rm s}$ frame, and momentum variables in the $\mathcal R_{\rm w}$ frame, as is common in cosmic ray physics~\cite{1967PhFl...10.2620H,*1968Icar....8...54D,*1972ApJ...172..319J,*1975MNRAS.172..557S, *1989ApJ...336..243S,*1993PhyU...36.1020B,*1993ApJ...405L..79W}, see in particular \cite{1985ApJ...296..319W,*1989ApJ...340.1112W,2018MNRAS.479.1747A,*2018MNRAS.479.1771A} for the relativistic regime. Appendix~\ref{app:fp} provides a detailed derivation of the following relativistic transport equation, which is based on the pioneering work of Webb~\cite{1985ApJ...296..319W,1989ApJ...340.1112W} and which properly incorporates the noninertial effects:
\begin{align}
 \gamma_{\rm w}&\left(\beta_{\rm w} {{p_{\vert\rm w}^t}}+{{p_{\vert\rm w}^x}}\right)\partial_x f_{\rm p} \nonumber\\
 & - \frac{{\rm d}u_{\rm w}}{{\rm d}x} \left(\beta_{\rm w}
 {{p_{\vert\rm w}^t}}+{{p_{\vert\rm w}^x}}\right){{p_{\vert\rm w}^t}}
 \left[{\mu_{\vert\rm w}}\partial_{{p_{\vert\rm w}}} +
   \frac{1-{\mu_{\vert\rm w}}^2}{{p_{\vert\rm w}}}\partial_{{\mu_{\vert\rm w}}}\right] f_{\rm p} \nonumber\\
& \quad\,=\,
 \frac{{{p_{\vert\rm w}^t}}}{2}\partial_{{\mu_{\vert\rm w}}}\left[\nu_{\vert\rm w}\left(1-{\mu_{\vert\rm w}}^2\right)\right]\partial_{{\mu_{\vert\rm w}}}
 f_{\rm p}\,.
 \label{eq:main1}
\end{align}
The second term on the l.h.s. describes the effective gravity associated to the deceleration of $\mathcal R_{\rm w}$.  The above equation has been implicitly averaged over the gyrophase $\phi_{\vert\rm w}$ of particles, so that the distribution function of the background plasma $f_{\rm p}$ depends solely on $x$, on ${p_{\vert\rm w}}$ and on ${\mu_{\vert\rm w}}\,\equiv\,{{p_{\vert\rm w}^x}}/{p_{\vert\rm w}}$.

As explained in App.~\ref{app:fp}, the scattering frequency $\nu_{\vert\rm w}$ depends a priori on both ${\mu_{\vert\rm w}}$ and ${p_{\vert\rm w}}$. In quasilinear theory, the collision term on the right hand side, hence $\nu_{\vert\rm w}$, would be evaluated to second order $\delta F^2$ in the random (Lorentz) force $\delta F$. However, quasilinear theory cannot be applied to describe the evolution of the background plasma, whose particles are mostly trapped in the Weibel filaments due to their relatively low momenta in $\mathcal{R}_{\rm w}$. Nevertheless, the above transport equation is expected to remain valid well beyond quasilinear theory because the collision operator is here dictated by symmetry considerations, which force it to be transverse to $\boldsymbol{{p_{\vert\rm w}}}$ in the $\mathcal R_{\rm w}$ frame. Here, we thus treat $\nu_{\vert\rm w}$ as a parameter and, to simplify the analysis, we neglect its dependence on ${\mu_{\vert\rm w}}$ and ${p_{\vert\rm w}}$. As will be discussed in Sec.~\ref{sec:nuw}, the standard scattering frequency for a marginally unbound particle, meaning with gyroradius $r_{\rm g\vert w}\,\gtrsim\,r_\perp$, where $r_\perp$ is the transverse size of the filament, provides a useful order of magnitude $\nu_{\vert\rm w}\,\sim\,\epsilon_B\,\omega_{\rm p}$.

For a given $4-$velocity profile $u_{\rm w}(x)$ characterizing the deceleration of $\mathcal R_{\rm w}$ in $\mathcal R_{\rm s}$, the following dimensionless parameter $\kappa$ emerges from Eq.~(\ref{eq:main1}):
\begin{equation}
  \kappa\,\equiv\,\nu_{\vert\rm w}^{-1}\frac{{\rm d}u_{\rm w}}{{\rm d}x}\,.
  \label{eq:defk}
\end{equation}
Where $\vert\kappa\vert\,\ll\,1$, the background plasma is effectively tied to the microturbulence, because it relaxes on a (shock frame) timescale $\gamma_{\rm w}{\nu_{\vert\rm w}}^{-1}$ that is smaller than the deceleration time scale of the turbulence, $\left\vert u_{\rm w}^{-1}{\rm d}u_{\rm w}/{\rm d}x\right\vert^{-1}$. Close to the shock front, however, it will be seen that $\vert\kappa\vert$ becomes of the order of  unity and larger, so that the background plasma decouples (temporarily) from this Weibel turbulence; coupling is eventually restored through relaxation on length scales $\nu_{\vert\rm w}^{-1}$ once the $\mathcal R_{\rm w}$ frame has reached its post-shock constant velocity. In this sense, $\vert\kappa\vert$ provides a useful order parameter to describe the shock transition. 

Equation~(\ref{eq:main1}) above can be rewritten as an infinite hierarchy of equations through a decomposition of $f_{\rm p}$ into Legendre polynomials:
\begin{equation}
  f_{\rm p}(x,{p_{\vert\rm w}},{\mu_{\vert\rm w}})\,=\,\sum_{n}f_n(x,{p_{\vert\rm w}})P_n({\mu_{\vert\rm w}})\,.
\end{equation}   
In the limit $\vert\kappa\vert\,\ll\,1$, the background plasma distribution function is approximately isotropic in the $\mathcal{R}_{\rm w}$ frame. We thus limit the Legendre expansion to its first two terms,
\begin{equation}
   f_{\rm p}\,\simeq\,f_{\rm p0}(x,{p_{\vert\rm w}}) + {\mu_{\vert\rm w}} f_{\rm p1}(x,{p_{\vert\rm w}})\,,
\label{eq:L01}
\end{equation}
with $\vert f_{\rm p1}\vert \,\ll\,f_{\rm p0}$ if $\vert\kappa\vert\,\ll\,1$, both functions being isotropic. In the above description, the magnitude of $f_{\rm p1}$ relative to $f_{\rm p0}$ characterizes the anisotropy/drift velocity of the background plasma in the $\mathcal R_{\rm w}$ frame. Hence, once $f_{\rm p1}$ becomes comparable to $f_{\rm p0}$, the decomposition in Eq.~(\ref{eq:L01}) becomes insufficient and should be extended to higher orders. We retain this description for the time being, for simplicity.

Taking the average of Eq.~(\ref{eq:main1}), respectively weighted by the first two Legendre polynomials, $P_0({\mu_{\vert\rm w}})\,=\,1$ and $P_1({\mu_{\vert\rm w}})\,=\,{\mu_{\vert\rm w}}$, we derive the system:
\begin{eqnarray}
  &&\gamma_{\rm w}\beta_{\rm w}{{p_{\vert\rm w}^t}}\partial_x f_{\rm p0} -
  \frac{1}{3}\frac{{\rm d}u_{\rm w}}{ {\rm d}x}{{p_{\vert\rm w}^t}}{p_{\vert\rm w}}\partial_{{p_{\vert\rm w}}}f_{\rm p0}
  \,=\, \nonumber\\
&&\quad\quad\quad -\frac{1}{3}\gamma_{\rm w}{p_{\vert\rm w}}\partial_x f_{\rm p1} +
  \frac{\beta_{\rm w}}{3}\frac{{\rm d}u_{\rm w}}{{\rm
      d}x}{{p_{\vert\rm w}^t}}^2\left(\partial_{{p_{\vert\rm w}}}f_{\rm p1} + \frac{2}{{p_{\vert\rm w}}}f_{\rm p1}\right)\,,\nonumber\\
&&  \label{eq:qlf0}\\ 
&&\frac{1}{3}\gamma_{\rm w}{p_{\vert\rm w}}\partial_x f_{\rm p0} - \frac{\beta_{\rm w}}{3}\frac{{\rm d}u_{\rm w}}{{\rm d}x}{{p_{\vert\rm w}^t}}^2\partial_{{p_{\vert\rm w}}}f_{\rm p0} \,=\,\nonumber\\
&&\quad\quad\quad -\frac{1}{3}\nu_{\vert\rm w}{{p_{\vert\rm w}^t}}f_{\rm p1}
  -\frac{1}{3}\gamma_{\rm w}\beta_{\rm w}{{p_{\vert\rm w}^t}}\partial_x f_{\rm p1}
\nonumber\\
&&\quad\quad\quad+
  \frac{1}{5}\frac{{\rm d}u_{\rm w}}{{\rm
      d}x}{{p_{\vert\rm w}^t}}{p_{\vert\rm w}}\left(\partial_{{p_{\vert\rm w}}}f_{\rm p1}+\frac{2}{3{p_{\vert\rm w}}}f_{\rm p1}\right)\,.\nonumber\\
  \label{eq:qlf1}
\end{eqnarray}
One can obtain a useful approximation to the solution to leading order in $\vert\kappa\vert$, by neglecting the second and third terms on the r.h.s of Eq.~(\ref{eq:qlf1}), which determines $f_{\rm p1}$ as
\begin{equation}
  f_{\rm p1}\,\simeq\, \kappa\beta_{\rm w} {{p_{\vert\rm w}^t}}\partial_{{p_{\vert\rm w}}}f_{\rm p0} - \frac{\gamma_{\rm w}}{\nu_{\vert\rm w}}
  \frac{{p_{\vert\rm w}}}{{{p_{\vert\rm w}^t}}}\partial_x f_{\rm p0}\quad\quad\left(\vert\kappa\vert\,\ll\,1\right)\,.
  \label{eq:qlf12}
\end{equation}
The smallness of $\vert\kappa\vert$ then validates our approximations.  Again,
to leading order in $\vert\kappa\vert$, one can replace the $\partial_x f_{\rm p0}$ in the above Eq.~(\ref{eq:qlf12}) equation by its value derived from Eq.~(\ref{eq:qlf0}), neglecting all terms of order $\kappa^2$ or higher. We thus end up with
\begin{equation}
  f_{\rm p1}\,\simeq\,\kappa\beta_{\rm w}\left(1-\frac{{{p_{\vert\rm w}}}^2}{3\beta_{\rm w}^2{{p_{\vert\rm w}^t}}^2}\right){{p_{\vert\rm w}^t}}\partial_{{p_{\vert\rm w}}}f_{\rm p0}
\quad\quad\left(\vert\kappa\vert\,\ll\,1\right) \,. \label{eq:qlf13}
\end{equation}
Inserting this relation in Eq.~(\ref{eq:qlf0}), one obtains to leading
order in $\vert\kappa\vert$ the Fokker-Planck equation for $f_{\rm p0}$:
\begin{align}
  \beta_{\rm w}\partial_x f_{\rm p0} -&  \frac{\beta_{\rm w}}{3u_{\rm w}} \frac{{\rm d}u_{\rm
        w}}{{\rm d}x}{p_{\vert\rm w}}\partial_{p_{\vert\rm w}}f_{\rm p0}\nonumber\\
&-\frac{1}{{{p_{\vert\rm w}}}^2}\partial_{{p_{\vert\rm w}}}\left[D_{{p_{\vert\rm w}}{p_{\vert\rm w}}}{{p_{\vert\rm w}}}^2\partial_{{p_{\vert\rm w}}}
f_{\rm p0}\right]\,=\,0\,.  \label{eq:fp1}
\end{align}
Some terms linear in $\partial_{{p_{\vert\rm w}}}$ have been neglected because they
renormalize the inertial term by small corrections of the order of $\vert\kappa\vert$ or higher. In agreement with our expansion to lowest order in $\kappa$, the above equation also neglects spatial diffusion, which would otherwise arise through the second term in the rhs of Eq.~(\ref{eq:qlf12}). The momentum space diffusion coefficient reads
\begin{equation}
  D_{{p_{\vert\rm w}}{p_{\vert\rm w}}}\,=\,\frac{\nu_{\vert\rm w}\beta_{\rm w}^2}{3\gamma_{\rm w}}\kappa^2\left(1-\frac{{{p_{\vert\rm w}}}^2}{3\beta_{\rm
        w}^2{{p_{\vert\rm w}^t}}^2}\right)^2{{p_{\vert\rm w}^t}}^2\,.
    \label{eq:dpp}
\end{equation}
The momentum diffusion term in Eq.~(\ref{eq:fp1}), which characterizes the stochastic heating due to the friction of particles on the microturbulence, scales with the square of the deceleration rate. The second term in this equation represents the heating induced by adiabatic plasma compression. This Fokker-Planck equation can be seen as a simplified one-dimensional version of the more general transport equation derived in~\cite{1985ApJ...296..319W,*1989ApJ...340.1112W,2018MNRAS.479.1747A,*2018MNRAS.479.1771A}. 
We also note that Ref.~\cite{1987PhR...154....1B} quotes in its equation~(3.46) a diffusion coefficient analogous to Eq.~(\ref{eq:dpp}) derived by G. F. Krimsky in the subrelativistic limit (the corresponding paper is not available).

At the microscopic level, dissipation results from stochastic acceleration of the background plasma particles in the sheared velocity flow that carries the turbulence at velocity $\beta_{\rm w}(x)$~\cite{L19}. The general relativistic transport equation given in~\cite{2018MNRAS.479.1747A,*2018MNRAS.479.1771A} also describes momentum diffusion driven by a shear term, which takes a form $\propto \left({\rm d}u_{\rm w}/{\rm d}x\right)^2$ as above, and an acceleration (or deceleration) term, which in the relativistic limit also takes the above form. Alternatively, this heating mechanism can be seen as some form of noninertial or differential first order Fermi acceleration: even though the turbulence is magneto-static in the $\mathcal R_{\rm w}$ frame, acceleration occurs because of the existence of an external force which keeps forcing the particles to interact with the turbulence, and because the velocity of this turbulence changes at every time step. The overall dissipative process can thus be pictured as a form of collisionless Joule heating, in which the effective gravity associated with the deceleration of the plasma plays the role of the driving electric field, while pitch angle scattering on the magnetostatic turbulence ensures momentum transfer. Finally, at the fluid level, the dissipative term becomes a form of viscosity.

\subsection{Moments}
As a result of the mixed coordinate system, the macroscopic quantities must be defined with care. In the lab frame $\mathcal R_{\rm s}$, the mean current density and energy-momentum tensors are defined as
\begin{eqnarray}
  {j_{\rm p}}^\alpha&\,=\,&2\pi\int{\rm d}{p_{\vert\rm w}}{\rm
    d}{\mu_{\vert\rm w}}\,\frac{{{p_{\vert\rm w}}}^2}{{{p_{\vert\rm w}^t}}}{{p_{\vert\rm w}}}^a{{e^\star}^\alpha}_a 
  f_{\rm p}\,,\nonumber\\ {T_{\rm p}}^{\alpha\beta}&\,=\,&2\pi\int{\rm d}{p_{\vert\rm w}}{\rm
    d}{\mu_{\vert\rm w}}\,\frac{{{p_{\vert\rm w}}}^2}{{{p_{\vert\rm w}^t}}}{{p_{\vert\rm w}}}^a{{p_{\vert\rm w}}}^b{{e^\star}^\alpha}_a {{e^\star}^
    \beta}_b  f_{\rm p}\,.\nonumber\\
&&  \label{eq:defT}
\end{eqnarray}
The tetrad ${{e^\star}^\alpha}_a$, which relates the $\mathcal R_{\rm w}$ to the $\mathcal R_{\rm s}$ frame, is defined in App.~\ref{app:fp}, see Eq.~(\ref{eq:etet}).  

The above formulae are general and valid to all orders in the expansion of $f_{\rm p}$ into Legendre polynomials. Restricting this development to its first two terms as above, we obtain
\begin{eqnarray}
  {j_{\rm p}}^t&\,=\,&\gamma_{\rm w}4\pi\int{\rm d}{p_{\vert\rm w}}\,\left[{{p_{\vert\rm w}}}^2 f_{\rm p0} + \frac{1}{3}\beta_{\rm w}\frac{{{p_{\vert\rm w}}}^3}{{{p_{\vert\rm w}^t}}}f_{\rm p1}\right]\,,\nonumber\\
  {j_{\rm p}}^x&\,=\,&\gamma_{\rm w}4\pi\int{\rm d}{p_{\vert\rm w}}\,\left[\beta_{\rm w}{{p_{\vert\rm w}}}^2 f_{\rm p0} + \frac{1}{3}\frac{{{p_{\vert\rm w}}}^3}{{{p_{\vert\rm w}^t}}}f_{\rm p1}\right]\,,\nonumber\\
  {T_{\rm p}}^{tx}&\,=\,&\gamma_{\rm w}^2\beta_{\rm w}4\pi\int {\rm d}{p_{\vert\rm w}}\,{{p_{\vert\rm w}}}^2{{p_{\vert\rm w}^t}}\left(1+\frac{{{p_{\vert\rm w}}}^2}{3{{p_{\vert\rm w}^t}}^2}\right)f_{\rm p0} \nonumber\\
&&\quad + \frac{1}{3}\gamma_{\rm w}^2\left(1+\beta_{\rm w}^2\right)4\pi\int{\rm d}{p_{\vert\rm w}}\,{{p_{\vert\rm w}}}^3f_{\rm p1}\,.
\label{eq:defT2}
\end{eqnarray}
The average drift velocity in the lab frame is ${j_{\rm p}}^x/{j_{\rm p}}^t$, which takes the form $(\beta_{\rm w}+\beta_{\rm p\vert w})/(1+\beta_{\rm
  w}\beta_{\rm p\vert w})$, with drift velocity $\beta_{\rm p\vert w}$ in the $\mathcal R_{\rm w}$ frame
\begin{equation}
  \beta_{\rm p\vert w}\,=\,\frac{1}{3}\frac{\int{\rm
      d}{p_{\vert\rm w}}\,{{p_{\vert\rm w}}}^3 f_{\rm p1}/{{p_{\vert\rm w}^t}}}{\int{\rm d}{p_{\vert\rm w}}\,{{p_{\vert\rm w}}}^2 f_{\rm p0}}\,.
  \label{eq:bdw}
\end{equation}
For a Maxwellian distribution characterized by a (proper) temperature $T_{\rm p}\,\ll\,m$, inserting Eq.~(\ref{eq:qlf13}) gives $\beta_{\rm p\vert w}\,\simeq\,-\kappa\beta_{\rm w}$. For a relativistically hot plasma, with $T_{\rm p}\,\gg\,m$, one  obtains $\beta_{\rm p\vert w}\,\simeq\,-\kappa\beta_{\rm w}\left[1-1/(3\beta_{\rm w}^2)\right]$.

Taking the moments of Eq.~(\ref{eq:quasil3}), we obtain the macroscopic equations of conservation of the current density and energy flux in the lab frame:
\begin{eqnarray}
  \frac{{\rm d}}{{\rm d}x} {j_{\rm p}}^x&\,=\,&0\,,\nonumber\\ \frac{{\rm d}}{{\rm
      d}x} {T_{\rm p}}^{tx}&\,=\,&-\frac{u_{\rm w}}{3}\nu_{\vert\rm w}4\pi\int{\rm d}{p_{\vert\rm w}}\,{{p_{\vert\rm w}}}^3 f_{\rm p1}\,.
  \label{eq:meq}
\end{eqnarray}
To obtain these equations, one must multiply Eq.~(\ref{eq:quasil2}) respectively by $1$ and ${{p_{\vert\rm w}}}^a {e^{\star}}^t_a$, then integrate over momentum space, paying attention to the spatial dependence of $\gamma_{\rm w}$. In the absence of scattering, the energy-momentum tensor of the background plasma is exactly conserved in the lab frame, since it is an inertial frame; a finite value of $\kappa$, however, implies a finite $f_{\rm p1}$ [see Eq.~(\ref{eq:qlf13}) or Eq.~(\ref{eq:bdw}], hence a source term for the energy flux evolution. 

To leading order in $\vert\kappa\vert$, one can make explicit the heating process by replacing $f_{\rm p1}$ with its expression given in terms of $f_{\rm p0}$ -- Eq.~(\ref{eq:qlf13}) -- in Eq.~(\ref{eq:meq}) above. The algebra is cumbersome but it simplifies considerably in either the
nonrelativistic $T_{\rm p}\,\ll\, m$ or ultrarelativistic $T_{\rm p}\,\gg\,m$ limits, as now detailed. 

In the nonrelativistic limit, $T_{\rm p}\,\ll\, m$, we split the energy-momentum tensor in its rest-mass and internal energy components, $p_{\rm p0}$ denoting the pressure associated with $f_{\rm p0}$. To lowest order in $p_{\rm p0}/(n_{\rm p0}m)$, one has
\begin{equation}
{T_{\rm p0}}^{tx}\,\simeq\,\gamma_{\rm w}m {j_{\rm p0}}^x + \frac{5}{2}\gamma_{\rm w}^2\beta_{\rm w} p_{\rm p0}\,.
\end{equation}
Then, the moment of the Fokker-Planck equation yields
\begin{align}
\beta_{\rm w}\frac{{\rm d}}{{\rm d}x} p_{\rm p0}\,+\,\frac{5}{3}\frac{\beta_{\rm w}}{u_{\rm w}}\frac{{\rm d}u_{\rm w}}{{\rm d}x}p_{\rm p0}\,-&\,\frac{2}{3}\frac{\kappa^2\beta_{\rm w}\nu_{\vert\rm w}}{\gamma_{\rm w}^2}{j_{\rm p0}}^x\,=\,0\,.
\end{align}
This equation describes both adiabatic heating through plasma compression (second term) and stochastic heating through turbulence-induced friction (third term). Assuming current conservation for $j_{\rm p0}^x$, which holds to lowest order in $\vert\kappa\vert$, the above can be rewritten for the temperature as
\begin{equation}
  \frac{{\rm d}}{{\rm d}x}\left(\vert u_{\rm w}\vert ^{2/3}\frac{T_{\rm p}}{m}\right)
  \,=\, \frac{2}{3}\frac{\kappa^2\beta_{\rm w}^2\nu_{\vert\rm w}}{\vert u_{\rm w}\vert^{1/3}}\,.
\label{eq:heat1}
\end{equation}
This compact form makes it clear that, in the absence of scattering the background plasma can only be adiabatically heated according to the law $\vert u_{\rm
  w}\vert^{2/3}T_{\rm p}/m\,=\,{\rm const.}$. This law properly describes the adiabatic compression of a 3D nonrelativistic gas along one spatial dimension. Dimensional analysis of Eq.~(\ref{eq:heat1}) further suggests that on a length scale of variation of the $4-$velocity, {\it i.e.} $\Delta x\,\simeq\,\vert u_{\rm w}/\kappa\vert$, the ratio of $T_{\rm p}/m$ varies by an amount of order $\vert\kappa\vert$; hence, the background plasma becomes relativistically hot once deceleration takes place and $\vert\kappa\vert$ approaches unity. Most of the heating is thus expected to occur in the shock transition.

Consider now the ultrarelativistic limit, $T_{\rm p}\,\gg\,m$, in which case the plasma pressure evolves according to
\begin{align}
\beta_{\rm w}\frac{{\rm d}}{{\rm d}x}p_{\rm p0} &\,+\,\frac{4}{3}\frac{\beta_{\rm w}}{u_{\rm w}}\frac{{\rm d}u_{\rm w}}{{\rm d}x}p_{\rm p0} \nonumber\\
&\,-\,\frac{4}{3}\left(1-\frac{1}{3\beta_{\rm w}^2}\right)^2\frac{\kappa^2\beta_{\rm w}^2\nu_{\vert\rm w}}{\gamma_{\rm w}}p_{\rm p0}\,=\,0\,.
\end{align}
One thus derives the equation for the temperature, assuming again current conservation to lowest order in $\vert\kappa\vert$,
\begin{equation}
\frac{{\rm d}}{{\rm d}x}\left(\vert u_{\rm w}\vert^{1/3} \frac{T_{\rm p}}{m}\right)\,=\, \frac{4}{3}\left(1-\frac{1}{3\beta_{\rm w}^2}\right)^2\frac{\kappa^2\beta_{\rm w}^2\nu_{\vert\rm w}}{\vert u_{\rm w}\vert^{2/3}} \frac{T_{\rm p}}{m}\,,
\label{eq:heat2}
\end{equation}
 which indicates that heating should become exponentially fast inside the shock transition, once $\vert\kappa\vert$ takes values of order unity and larger.

\subsection{Numerical evaluation}
The previous analytical approximations, obtained in the limit $\vert\kappa\vert\,\ll\,1$, offer useful insight into the nature and the efficiency of the heating process, but they cannot cover the whole length of the precursor, in particular the near-shock region. For a more accurate description, we here present the results of a numerical integration of the dynamics represented by Eq.~(\ref{eq:main1}). 

One way to proceed is to decompose $f_{\rm p}$ in Legendre polynomials up to some order $L$, as in Eq.~(\ref{eq:L01}) for $L=1$, then integrate numerically the analog system of partial differential equations in $p$ and $x$, or to evaluate the moments of these equations to obtain a hierarchy of fluid-like first order differential equations in $x$. However, the convergence of the decomposition in Legendre polynomials is usually slow, so that it becomes necessary to push this decomposition to high $L$ in order to solve the dynamics close to the shock front. Indeed, the closer to the shock front, the stronger the anisotropy of $f_{\rm p}$ in the $\mathcal R_{\rm w}$ frame, due to the decoupling of the plasma from the Weibel turbulence. In practical terms, we find that it is possible to follow the dynamics of the background plasma up to $x\,\approx\, 10^2\,c/\omega_{\rm p}$ with $L\,=\,2$, with a numerical integration that is not devoid of instabilities in $p-$space.

We thus follow a different approach here, making use of Monte Carlo integration of particle trajectories subject to pitch-angle diffusion in a noninertial frame. Our parallelized solver stochastically propagates a large amount of particles in a box of size close to that of the shock precursor. As above, the deceleration law is characterized by $u_{\rm w}(x)$. Each particle is subject, in the $\mathcal R_{\rm w}$ frame, to an It\^o-type stochastic equation 
\begin{eqnarray}
  {\rm d}{\mu_{\vert\rm w}} &\,=\,& \sqrt{2\nu_{\vert\rm w}{\rm d}t_{\vert\rm w}}\xi\,,\nonumber\\
  {\rm d}{p_{\vert\rm w}^x} &\,=\,& {p_{\vert\rm w}}\,{\rm d}{\mu_{\vert\rm w}}  - \left(\beta_{\rm w}
  {{p_{\vert\rm w}^t}}+{{p_{\vert\rm w}^x}}\right)\frac{{\rm d}u_{\rm w}}{{\rm d}x}{\rm d}t_{\vert\rm w}\,,\nonumber\\
    \label{eq:numstoch}
\end{eqnarray}
where $\xi\,\sim\,\mathcal N(0,1)$ is a normally distributed random
number representing white noise, and at all times, the transverse momentum is set according to  $p_{\perp\vert\rm w}\,=\,\sqrt{1-{\mu_{\vert\rm w}}^2}{p_{\vert\rm w}}$. 

In the absence of the inertial term, {\it i.e.}, the second term on the r.h.s. of the second equation, the norm ${p_{\vert\rm w}}$ is a constant and no heating occurs. The time step is taken to be constant in the shock-front frame. The equations of motion, however, are solved in the noninertial $\mathcal{R}_{\rm w}$ frame, using the time step 
\begin{equation}
{\rm d}t_{\vert\rm w} = \frac{{p_{\vert\rm w}^t}\,{\rm d}t}{\gamma_{\rm w} \left(\beta_{\rm w}  {p_{\vert\rm w}^x} + {p_{\vert\rm w}^t} \right) } \,,
\end{equation}	
with, in practice, ${\rm d}t = 0.01 \omega_{\rm p}^{-1}$. Particles are injected far from the shock, isotropically in $\mathcal R_{\rm w}$ with a temperature $T_\infty\,=\,0.01\,m$, in acccordance with that used in our PIC simulations. 

This Monte Carlo model is characterized by two parameters: $\nu_{\vert\rm w}$ and $u_{\rm w}(x)$. We retain $\nu_{\vert\rm w}$ as a free parameter, but fix the deceleration law $u_{\rm w}(x)$ according to one of our theoretical models developed in Paper~I~\cite{pap1} and validated on our large-scale PIC simulations. This model describes the nonlinear phase of the CFI as a local quasistatic pressure equilibrium between the particle populations and the magnetic field in $\mathcal R_{\rm w}$, borrowing on the study of Ref.~\cite{Vanthieghem_2018}. This approximation implies that the physical conditions evolve slowly enough throughout the precursor, so that the system is close to steady state at every location. Detailed comparisons between this model and our reference PIC simulations show that it is able to capture, to within a factor of two, the (subrelativistic) relative velocity $\beta_{\rm w\vert p}$. The simulations provide the energy density, mean drift $4-$velocity and temperature of the various species which are used as inputs of the model. 

As discussed in Sec.~\ref{sec:declaw}, our simulations distinguish background plasma from suprathermal particles acccording to the sign of their $x-$momentum and the number of reversals they have experienced. This distinction becomes irrelevant within the last skin depths to the shock front because the background plasma particles then suffer strong deflections. Hence, the quantities $u_{\rm p\vert d}$ and $T_{\rm p}$ extracted from the simulation become inaccurate there, and so does $\beta_{\rm w\vert p}$ by implication. Therefore, we complement the profile of $u_{\rm w}(x)$ with a plateau once $u_{\rm w}(x)$ reaches the shock crossing value $-1/\sqrt{3}$ (for a 2D3V simulation). This plateau extends over to negative $x$ values and is not visible in the following figures because of the choice of a logarithmic scale in $x$.

\begin{figure}
\includegraphics[width=0.45\textwidth]{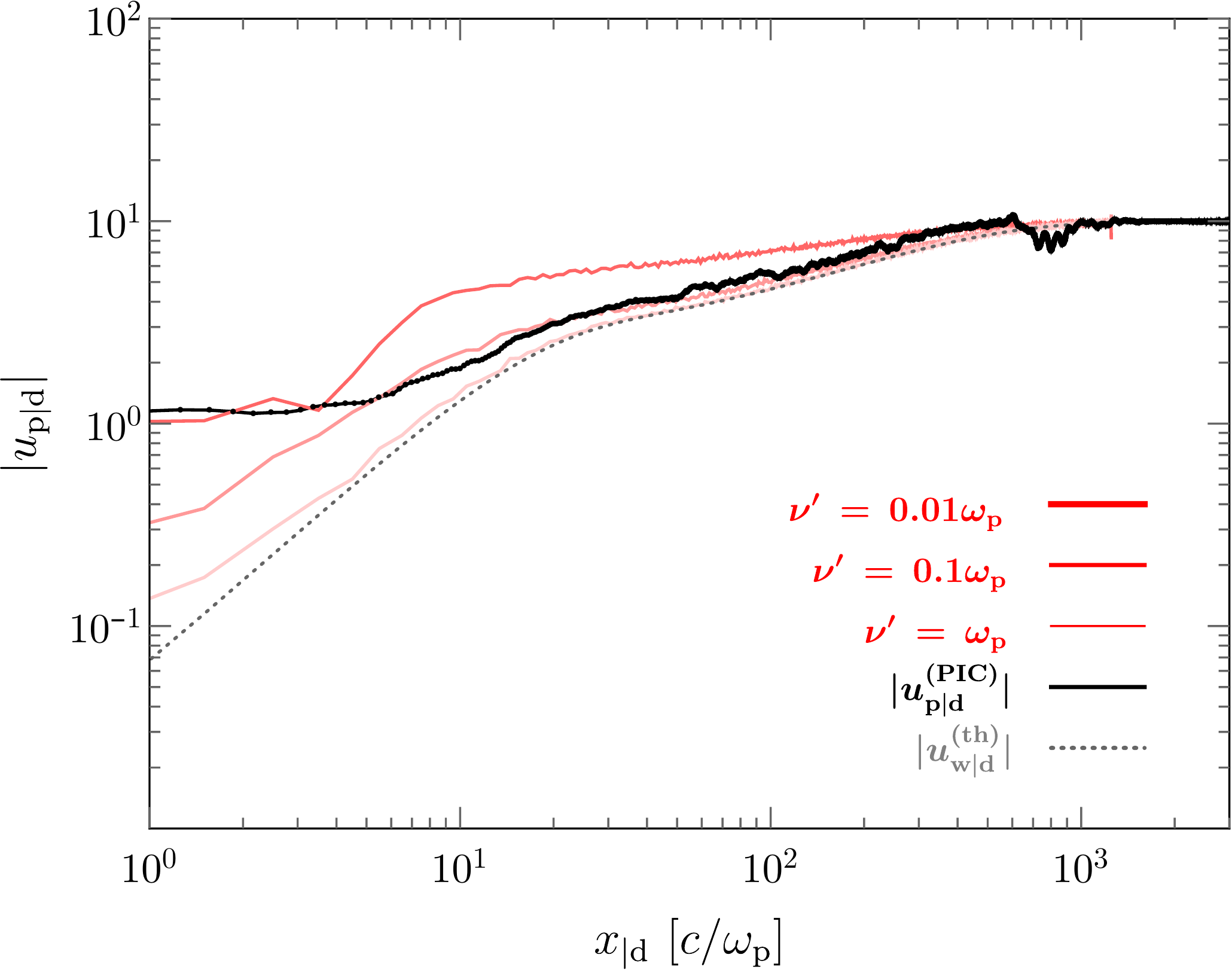}\smallskip\\
\includegraphics[width=0.45\textwidth]{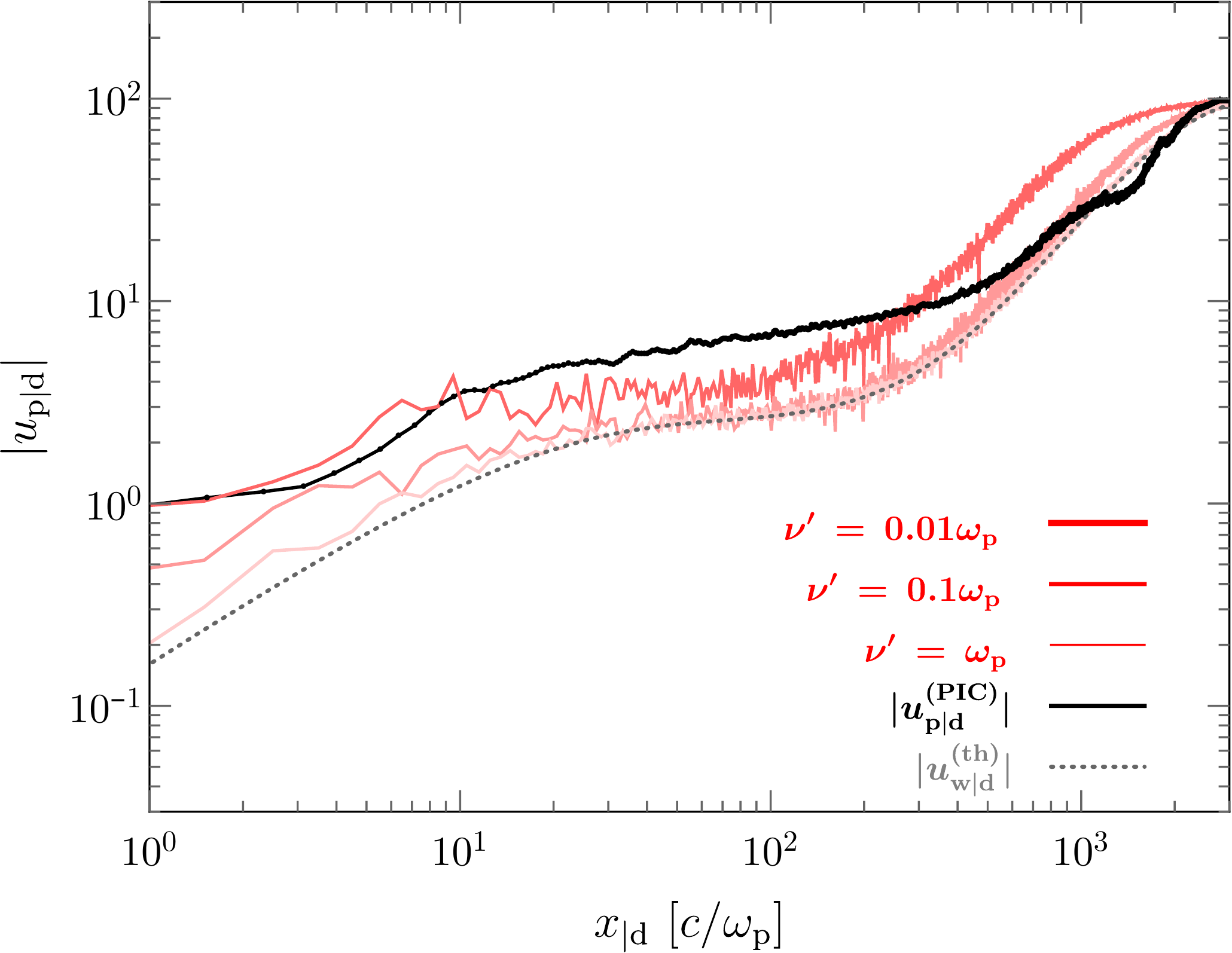}
  \caption{Profiles of the background plasma $4-$velocity $\vert u_{\rm p\vert d}\vert$ in the simulation frame $\mathcal R_{\rm d}$ as a function of distance to shock $x$. In red, the results of the Monte Carlo computation, for different values of the scattering frequency (light red: $\nu_{\vert\rm w}\,=\,\omega_{\rm p}$, medium red: $\nu_{\vert\rm w}\,=\,0.1\omega_{\rm p}$ and dark red: $\nu_{\vert\rm w}\,=\,0.01\omega_{\rm p}$); in black, the value measured in the PIC simulation. We stress that the values extracted from the PIC simulation at $x\lesssim10c/\omega_{\rm p}$ are inaccurate because the distinction between background plasma particles and shock-heated particles becomes difficult. The dotted gray line shows the 4-velocity $\vert u_{\rm w} \vert$, inferred from the PIC simulation, and used as input in the Monte Carlo computation.  Top panel: $\gamma_{\infty\vert\rm d}\,=\,10$ corresponding to $\gamma_{\infty}\,=\,17$; bottom panel: $\gamma_{\infty\vert\rm d}\,=\,100$ corresponding to $\gamma_{\infty}\,=\,173$.
  \label{fig:uvsnu} }
\end{figure}

\begin{figure}
\includegraphics[width=0.45\textwidth]{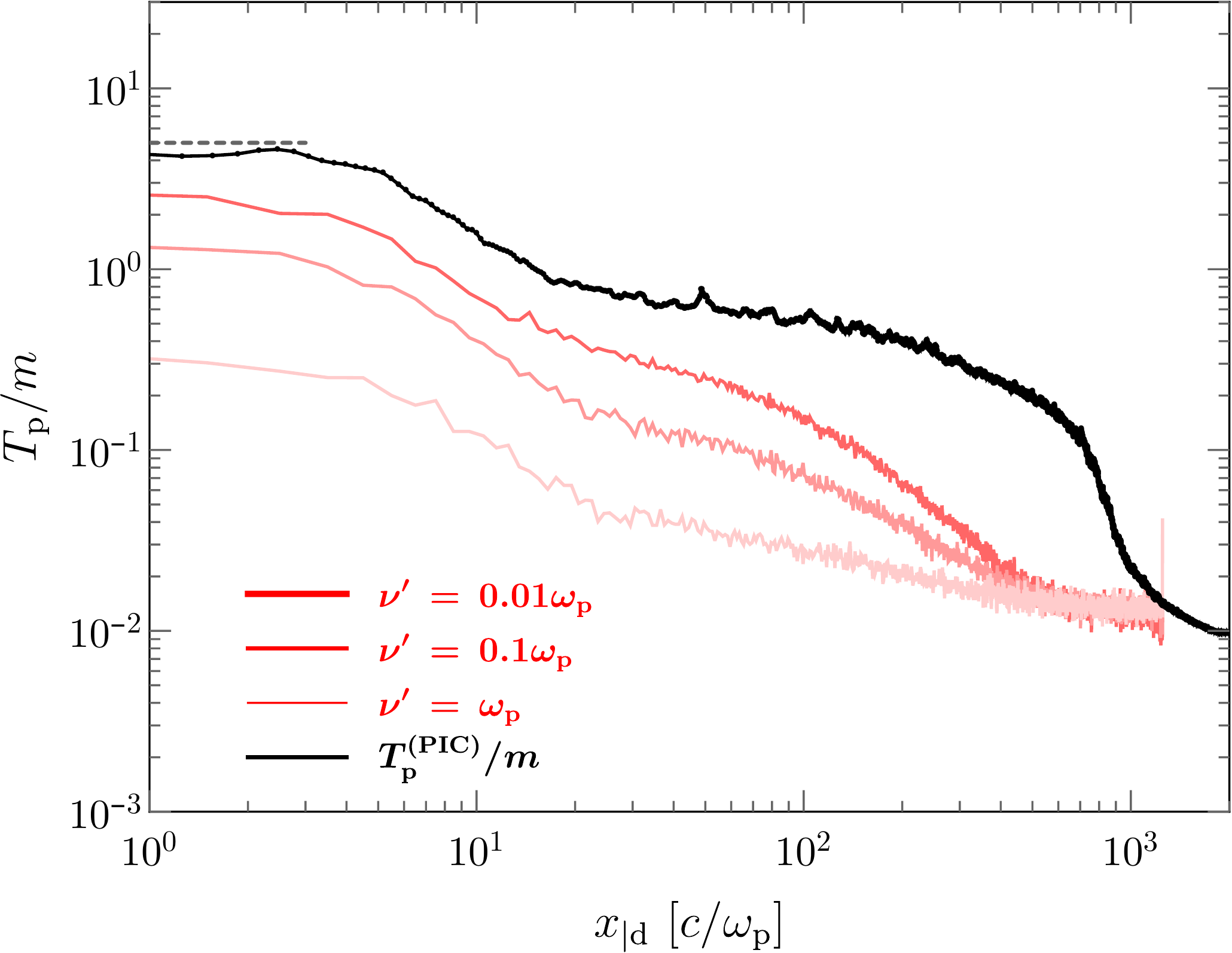}\smallskip\\
\includegraphics[width=0.45\textwidth]{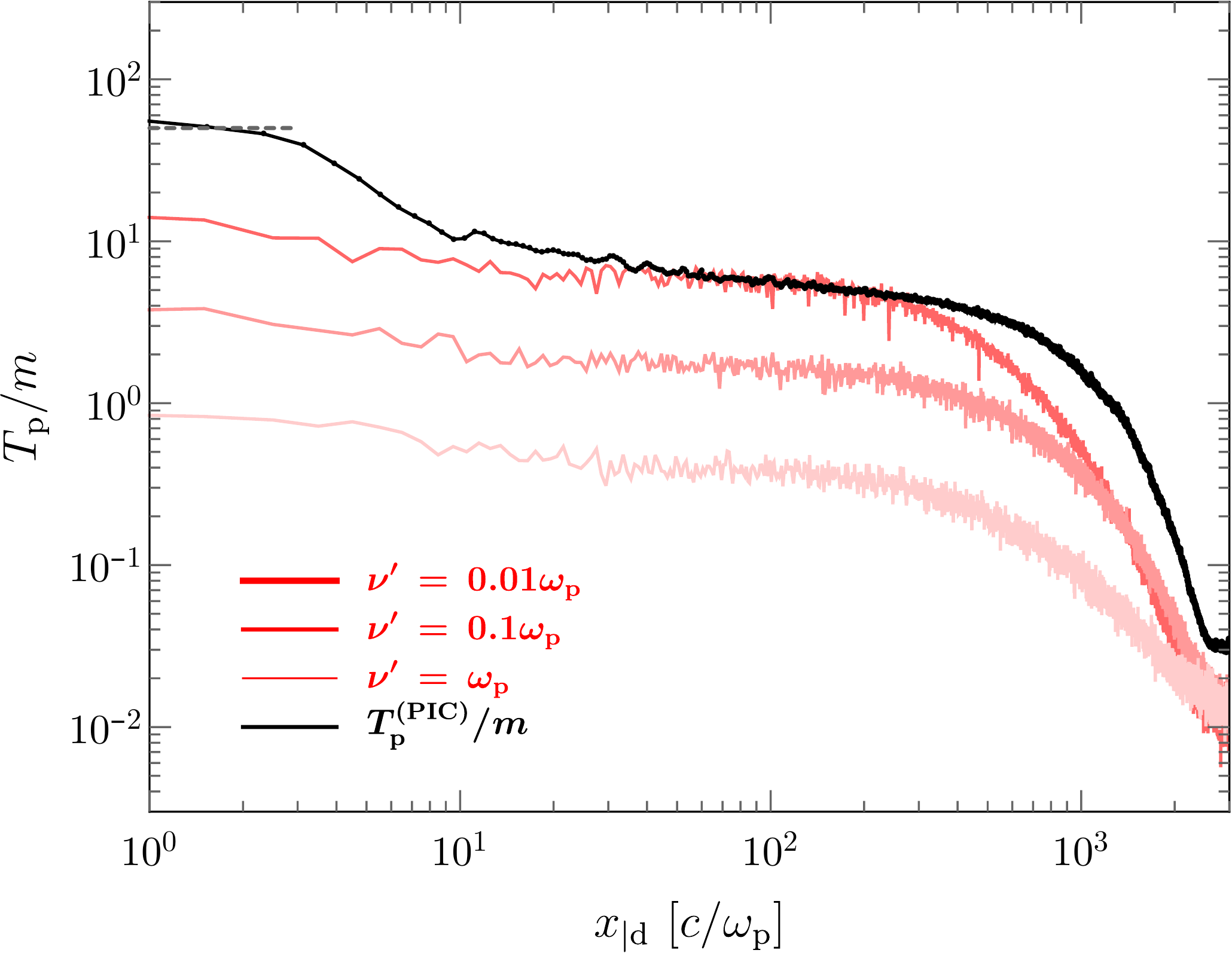}
  \caption{Same as Fig.~\ref{fig:uvsnu} for the profiles of the temperature $T_{\rm p}$ as a function of distance to shock $x$, in the simulation frame. We stress that the values extracted from the PIC simulation at $x\lesssim10c/\omega_{\rm p}$ are inaccurate because the distinction between background plasma particles and shock-heated particles becomes difficult. The dashed gray line  indicates the expected final values of $T_{\rm p}$ corresponding to the fluid shock crossing conditions for a relativistic unmagnetized shock in 2D. 
  \label{fig:Tvsnu} }
\end{figure}

\begin{figure}
\includegraphics[width=0.45\textwidth]{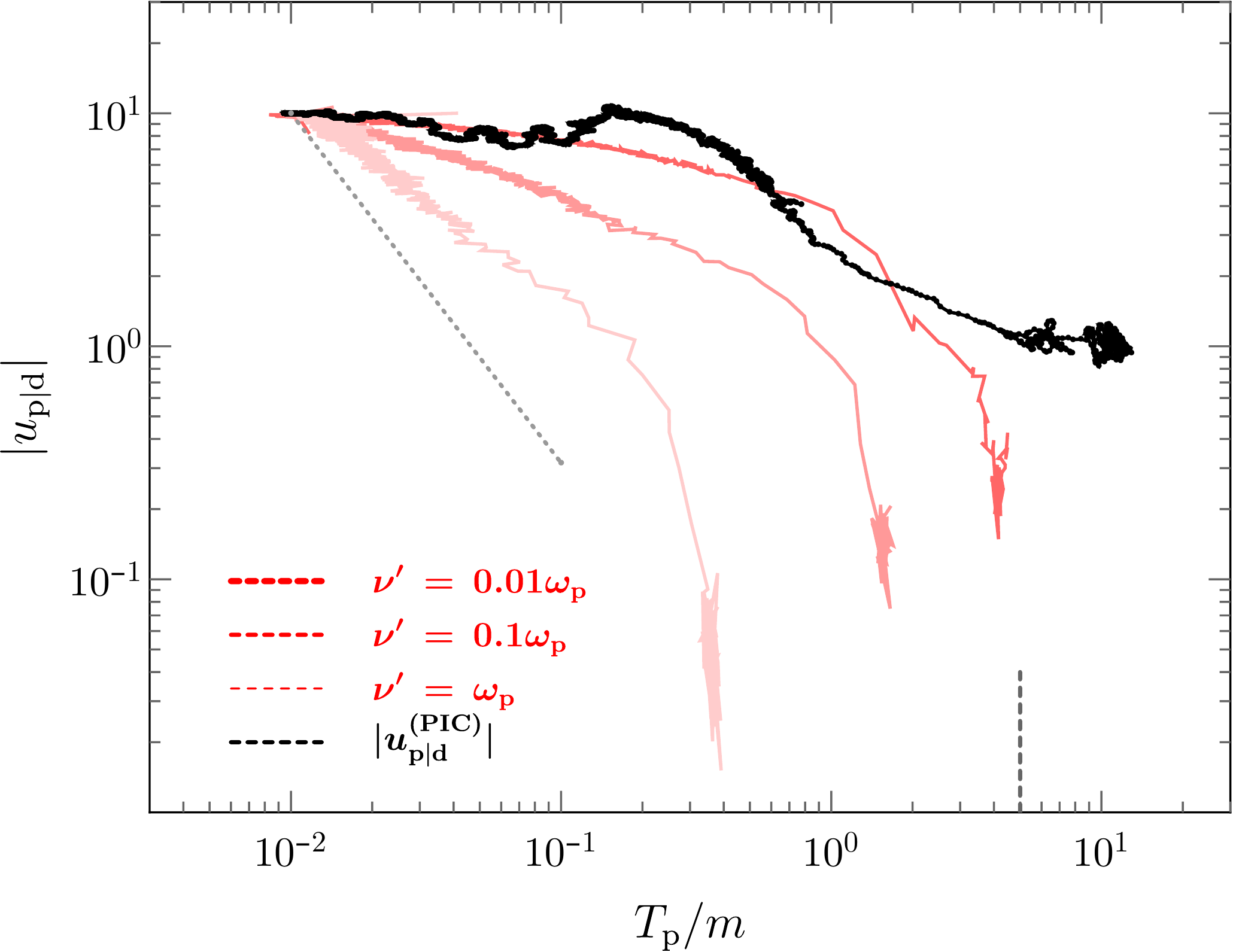}\smallskip\\
\includegraphics[width=0.45\textwidth]{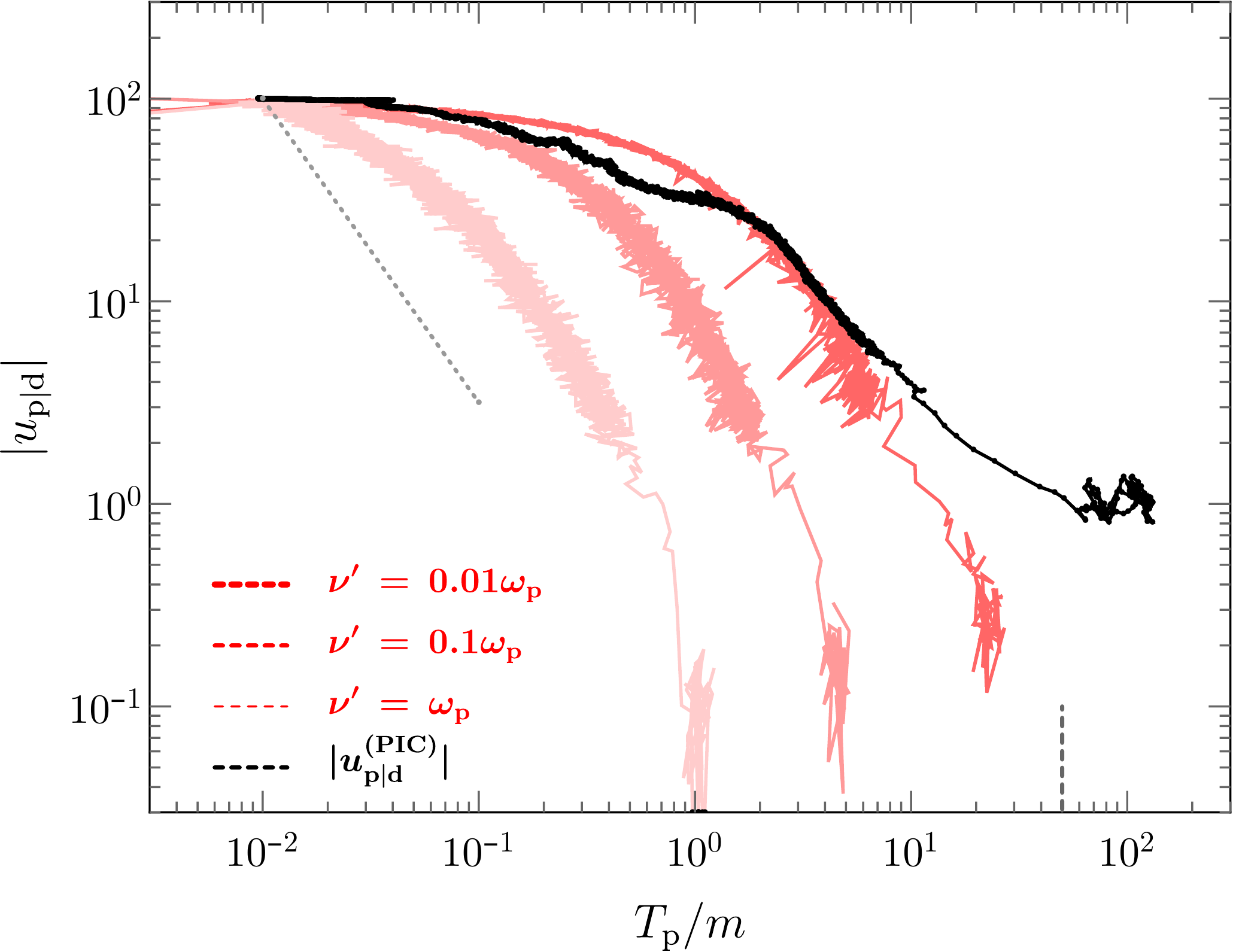}
  \caption{Same as Fig.~\ref{fig:uvsnu}, but now representing the trajectory of the background plasma in the plane $(T_{\rm p},\,\,\vert u_{\rm p}\vert)$. We stress that the values extracted from the PIC simulation at large $T_{\rm p}$ are inaccurate because the distinction between background plasma particles and shock-heated particles becomes difficult. The dashed gray line indicates the expected final values of $T_{\rm p}$ corresponding to the fluid shock crossing conditions for a relativistic unmagnetized shock in 2D (the corresponding value of $u_{\rm p}^x$ is zero in the simulation frame). The dotted line shows the adiabatic law of compression for a non-relativistic gas, $T_{\rm p}\,\propto\vert u_{\rm p}^x\vert^{-2/3}$.
  \label{fig:uvsT} }
\end{figure}

Because our Monte Carlo model is defined in the shock frame,  its results must be Lorentz transformed to the downstream frame in which the PIC simulations are run. We should also point out that the present model assumes a steady-state situation, while PIC simulations are by definition time-dependent; therefore the comparison between the two turns out to be slightly delicate. For one, we provide this comparison at a fixed (simulation frame) time $t_{\rm\vert d}$, not at a fixed $\mathcal R_{\rm s}$-frame time. Moreover, in our PIC simulations, the precursor enlarges in proportion to $t_{\rm\vert d}$ -- the time of the simulation -- while the maximal energy of accelerated particles inside this precursor increases roughly with $\sqrt{t_{\rm\vert d}}$, so that their diffusion length increases as $t_{\rm\vert d}$. As discussed in Paper III~\cite{pap3}, one may thus consider that, at all times, a portion of the precursor is in steady-state, because the scattering time of the particles is smaller than $t_{\rm\vert d}$, while the rest of the precursor is populated by particles with an energy so large that they have not yet scattered significantly. Since the typical energy of particles increases with distance to the shock, as a result of the smaller scattering length of lower energy particles, the near precursor, close to the shock front, is expected to be in steady state, while the far precursor should be described with a time-dependent model. In the $\gamma_\infty\,=\,17$ (resp. $\gamma_\infty\,=\,173$) simulation that we use as a testbed here, the steady-state approximation seems to be valid for $x\,\lesssim\, 10^3\omega_{\rm p}^{-1}$ (resp. $x\,\lesssim\, 2\times10^3\omega_{\rm p}^{-1}$), while the precursor extends up to $2\times 10^3\omega_{\rm p}^{-1}$ (resp. $3.5\times 10^3\omega_{\rm p}^{-1}$).

We thus use our Monte Carlo code with the above $u_{\rm w}^{\rm (th)}(x)$ (rather, an interpolation of it) as input for the deceleration law to obtain theoretical predictions for $u_{\rm p\vert d}(x)$ and $T_{\rm p}(x)$, which we compare with the values measured in PIC simulations in Figs.~\ref{fig:uvsnu}, \ref{fig:Tvsnu} and \ref{fig:uvsT}. We consider three representative values for our scattering frequency parameter: $\nu_{\vert\rm w}\,=\,1,\,0.1,\,0.01\,\omega_{\rm p}$ represented respectively in light, medium and dark red in those figures. In each of these figures, the left (resp. right) panel provides the comparison for $\gamma_\infty\,=\,17$ (resp. $\gamma_\infty\,=\,173$). 

 Figure~\ref{fig:uvsnu} presents the 4-velocity profile of the background plasma as a function of distance to the shock. As the scattering frequency increases, $u_{\rm p\vert d}$ lies closer and closer to the input theoretical profile of $u_{\rm w\vert d}$, as one would expect, because a large scattering frequency implies a stronger coupling of the plasma to the turbulence. Although the Monte Carlo curves do not account for the detailed evolution of the measured background plasma 4-velocity, they provide a reasonable match to the overall PIC profile. We emphasize that our parameter $\nu_{\vert\rm w}$ is constant, understood as an average quantity over the precursor, whereas one should rather expect the exact $\nu_{\vert\rm w}$ to depend on $x$, just like other quantities. At $x\,\lesssim\,1\,c/\omega_{\rm p}$ (including negative values not visible in these figures), the profile $u_{\rm w\vert d}$ relaxes further to subrelativistic values, and so does the background plasma on length scales of tens of $c/\omega_{\rm p}$, until it eventually matches the asymptotic $u_{\rm w/vert d}$, with a temperature a factor of $\sim\,2$ or so larger than the Monte Carlo curves shown in Fig.~\ref{fig:Tvsnu}. The Monte Carlo calculations indicate that larger scattering frequencies lead to smaller post-shock temperatures. This can be understood as follows: as $\nu_{\vert\rm w}\ell_{\rm prec}\,\rightarrow\,+\infty$, the background plasma behaves as an ideal fluid, therefore it obeys the adiabatic compression law, see Eqs.~(\ref{eq:heat1}), (\ref{eq:heat2}). This cannot suffice of course, as the large entropy jump at the shock requires a significant amount of dissipation, which scales in inverse proportion to $\nu_{\rm w}$, see Eqs.~(\ref{eq:heat1}), (\ref{eq:heat2}) in the small $\vert\kappa\vert$ limit. A value $\nu_{\vert\rm w}\,\sim\,0.01\omega_{\rm p}$ seems to account reasonably well for the shock jump condition for both values of $\gamma_\infty$, see Fig.~\ref{fig:uvsT}, which combines the above profiles in a trajectory of the background plasma in the plane $\vert u_{\rm p\vert d}\vert$ vs $T_{\rm p}$. The black curves plot the trajectories extracted from the PIC simulations, while the red curves correspond to the Monte Carlo simulations with different values of $\nu_{\vert\rm w}$. The dotted line shows the trajectory that would be expect for pure adiabatic compression of a subrelativistic 2D gas, $T_{\rm p}\,\propto\,\vert u_{\rm p}\vert^{-2/3}$. As observed above, the trajectories corresponding to larger values of $\nu_{\vert\rm w}$ remain closer to these adiabatic trajectories, and would provide an exact match in the limit $\nu_{\vert\rm w}\ell_{\rm prec}\,\rightarrow\,+\infty$. These plots show a rather nice agreement of the overall trajectories for our best-fit scattering frequency $\nu_{\vert\rm w}\,=\,0.01\omega_{\rm p}$.

The left panel of Fig.~\ref{fig:Tvsnu} for $\gamma_\infty\,=\,17$, however, reveals a mismatch between the measured temperature and that modeled at distances $10^2c/\omega_{\rm p}\,\lesssim\,x\,\lesssim\,10^3\,c/\omega_{\rm p}$. We interpret this as a transient effect, which appears in this simulation of duration $3600\omega_{\rm p}^{-1}$, but that fades on longer timescales; this effect, in particular, becomes substantially milder in the simulation $\gamma_\infty\,=\,173$ of duration $6900\,\omega_{\rm p}^{-1}$ and it is absent of our longest simulation of duration $10700\,\omega_{\rm p}^{-1}$ for $\gamma_\infty\,=\,17$\footnote{We do not use this simulation to benchmark our model because its diagnostics cannot disentangle the suprathermal and background plasma particles finely enough for our present purposes.}. This sudden heating of the background plasma from $T_{\rm p}\,\simeq\,10^{-2}m$ to $T_{\rm p}\,\simeq\,10^{-1}m$ over a few hundreds of $c/\omega_{\rm p}$ can be related to the oscillatory slowdown feature in the plasma 4-velocity at $x\,\simeq\,600-1000\,c/\omega_{\rm p}$ (see Fig.~\ref{fig:uvsnu}), and more particularly to the change in the momentum distribution of suprathermal particles: at $x\,\gtrsim\,10^3\,c/\omega_{\rm p}$, the suprathermal plasma is dominated by the particles that were specularly reflected off the mirror at early times, while at $x\,\lesssim\,10^3\,c/\omega_{\rm p}$, the large inertia particles that result from acceleration on the shock front form the bulk of this population. We expect that on longer timescales, this preheating effect would fade away, as for other simulations, and that the theoretical $T_{\rm p}(x)$ curve would better match the PIC simulations.

\subsection{Discussion}

\subsubsection{The scattering frequency of the background plasma}\label{sec:nuw}
Our model indicates a best-fit value of $\nu_{\vert\rm w}\,=\,0.01\omega_{\rm p}$ for the effective scattering frequency of the background plasma in the $\mathcal R_{\rm w}$ frame. 

As discussed in Paper~I~\cite{pap1}, the degree of nonlinearity of the filamentation of the background plasma in $\mathcal R_{\rm w}$ can be quantified using the parameter $\Xi_{\rm p\vert w}\,\simeq\,e \vert\beta_{\rm p\vert w}\vert \delta A^x/T_{\rm p}$, with $\delta A^x$ the $x-$component of the electromagnetic potential four-vector. In our PIC simulations, this nonlinearity parameter is smaller than unity far in the precursor, but close to unity in the near precursor, indicating that a significant fraction of background plasma particles are trapped in the filaments, all the more so within hundreds of skin depths of the shock front. Individual particles are trapped in a filament if their gyroradius $r_{\rm g\vert w}\,<\,r_\perp$, with $r_\perp\,\sim\,10\,\omega_{\rm p}^{-1}$ the filament transverse radius. Defining $\overline \gamma_{\rm p\vert\rm w}$ the typical Lorentz factor of a background plasma particle in $\mathcal R_{\rm w}$, {\it i.e.}, $\overline\gamma_{\rm p\vert\rm w}\,\simeq\,{\rm max}\left[1,T_{\rm p}/m\right]$ (neglecting the bulk drift velocity in front of the thermal velocity in $\mathcal R_{\rm w}$), we find that most particles are trapped if $T_{\rm p}/m\,<\,\left(r_\perp\omega_{\rm p}\right)^2 \epsilon_B$ (for $T_{\rm p}\,<\,m$), or $T_{\rm p}/m\,<\,\left(r_\perp\omega_{\rm p}\right)\epsilon_B^{1/2}$ ($T_{\rm p}\,>\,m$).

For those trapped particles, one can form an estimate of the scattering frequency as follows. Assume that trapped particles execute oscillating orbits inside the filaments, bouncing on the magnetic field barrier that reaches a peak at $r_\perp$. The betatron frequency characterizing those oscillations is $\omega_{\beta\vert\rm w}\,=\,\sqrt{\pi/2}\left(r_\perp r_{\rm g\vert w}/\overline\beta_{x\vert\rm w}\right)^{-1/2}$, with $\overline\beta_{x\vert\rm w}$ the $x-$velocity of the particle. In the following, we assume $\overline\beta_{x\vert\rm w}\,\sim\,1$, corresponding to $T_{\rm p}\,\gtrsim\, m$, as observed in the near precursor. At each rebound, the particle is deflected by an angle $\Delta \alpha_{\vert\rm w}\,\sim\,\pm r_\perp\omega_{\beta\vert\rm w}$. However, the force remains coherent over the trajectory of the particle inside the filament, to decohere only once the particle has exited the filament over a parallel coherence length $\ell_\parallel$, on a timescale $\Delta t_{\rm\vert w}\,\sim\,\ell_{\parallel\vert\rm w}$. Therefore, the angular diffusion frequency can be written
\begin{equation}
{\nu_{\vert\rm w}}^{\rm (trapped)}\,=\,\frac{\Delta\alpha_{\rm\vert w}^2}{\Delta t_{\rm\vert w}}\,\sim\,
\frac{r_\perp}{\ell_{\parallel\vert w}}\,\epsilon_B^{1/2}\,{\overline \gamma_{\rm p\vert\rm w}}^{-1}\omega_{\rm p}\,,
\label{eq:nutr}
\end{equation}
where $\overline \gamma_{\rm p\vert\rm w}$ represents the typical Lorentz factor of background plasma particles in $\mathcal R_{\rm w}$. For typical values of $r_\perp/\ell_{\parallel\vert\rm w}\,\sim\,0.1$, $\epsilon_B\,\sim\,0.01$ and $\overline \gamma_{\rm p\vert\rm w}\,\sim\,1$, one obtains the same order of magnitude as the best-fit value, $\nu_{\vert\rm w}\,\sim\,0.01\,\omega_{\rm p}$.

Consider now the untrapped population, with gyroradius $r_{\rm g\vert w}\,\gtrsim\,r_\perp$. In this case, the standard estimate of the scattering frequency leads to $\nu_{\vert\rm w}\,\sim\,r_\perp/r_{\rm g\vert w}^2$, {\it i.e.},
\begin{equation}
{\nu_{\vert\rm w}}^{\rm (untrapped)}\,\sim\,\left(r_\perp\omega_{\rm p}\right)\,\epsilon_B\,{\overline \gamma_{\rm p\vert\rm w}}^{-2}\omega_{\rm p}\,,
\label{eq:nuuntr}
\end{equation}
which also provides a reasonable order of magnitude for $r_\perp\omega_{\rm p}\,\sim\,1-10$, $\epsilon_B\,\sim\,0.01$ and $\overline \gamma_{\rm p\vert\rm w}\,\sim\,1$.

\begin{figure}
\includegraphics[width=0.45\textwidth]{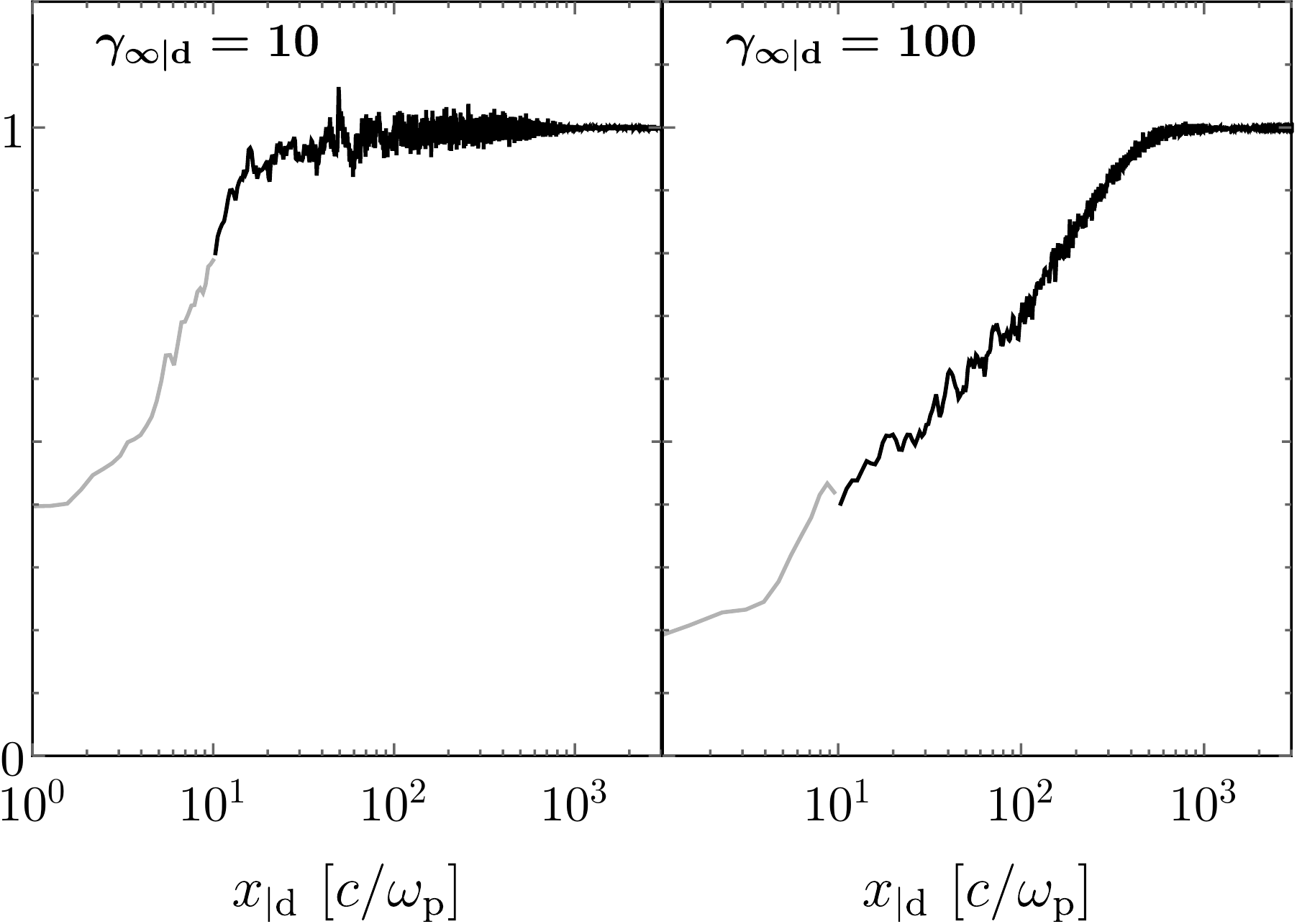}
 \caption{Spatial profile of $\gamma_{\rm p\vert d}\beta_{\rm p\vert d}n_{\rm p}/(\gamma_{\infty\vert\rm d}\beta_{\infty\vert\rm d}n_\infty)$ extracted from our two reference PIC simulations, as indicated. This quantity decreases towards the shock as a result of a finite scattering frequency, because once the momentum of a particle has changed sign, this particle is no longer characterized in this post-processing as a background plasma particle. This figure shows that about half of the particles are scattered at least once over the crossing of the precursor.
  \label{fig:cdensp} }
\end{figure}

The average scattering frequency of the background plasma can also be directly estimated from our PIC simulations by the following counting argument. In the PIC simulations, the background plasma particles are defined as those particles that propagate towards the $-\bm{x}$ direction and that have not undergone any turn-around. If the scattering frequency were exactly zero, then the current density $u_{\rm p}^{\rm (PIC)}n_{\rm p}^{\rm (PIC)}$ -- the superscript (PIC) referring to how these particles are defined in the PIC simulation -- would be exactly conserved. At finite scattering frequency, however, this current density must decrease at a spatial rate $\nu_{\vert\rm w}/\gamma_{\rm w\vert d}$ along the plasma trajectory in the simulation frame. Figure~\ref{fig:cdensp} shows the spatial profile of this current density for our two reference PIC simulations, indicating that indeed, over the crossing of the precursor, about half of the background plasma particles have experienced at least one turnaround. One can then approximate the scattering frequency with $\nu_{\vert\rm w}\,\sim\,\langle\gamma_{\rm w\vert d}\rangle/\ell_{\rm prec\vert d}$. For $\gamma_{\infty\vert\rm d}\,=\,10$, we estimate an average $\langle\gamma_{\rm w\vert d}\rangle\,\sim\,10$ over the precursor length scale $\ell_{\rm prec\vert\rm d}\,\sim\,10^3\omega_{\rm p}^{-1}$, see Figs.~1 and 2 in Paper~I~\cite{pap1}, giving $\nu_{\vert\rm w}\,\sim\,0.01\omega_{\rm p}$. For $\gamma_{\infty\vert\rm d}\,=\,100$, we estimate an average $\langle\gamma_{\rm w\vert d}\rangle\,\sim\,30$ over the precursor length scale $\ell_{\rm prec\vert\rm d}\,\sim\,2\times10^3\omega_{\rm p}^{-1}$, see Figs.~1 and 2 in Paper~I~\cite{pap1} and~\cite{L1}, giving $\nu_{\vert\rm w}\,\sim\,0.015\omega_{\rm p}$, both in good agreement with the theoretically inferred value.

\subsubsection{The possible role of longitudinal electric fields}\label{sec:dEx}

The above microscopic model of the heating and the deceleration of the background plasma neglects the possible contribution of longitudinal electric fields. Accordingly, we do not expect any coherent electrostatic component $\delta E_x$ to arise in the present case of a pair shock. One does, however, expect an inductive $\delta E_x$ associated to growth of the CFI in the precursor and an electrostatic $\delta E_x$ associated to the broadband nature of the CFI. Neither of these would be coherent, since the inductive component should be modulated along the $y-$direction as $\delta B_z$, while the electrostatic component should be modulated at least along $x$ with reversal scale $\sim\,k_x^{-1}$. 

As already mentioned, the energy density of the longitudinal electric fields that we observe in the PIC simulations is significantly smaller than that of the magnetic field, even when boosting back to the $\mathcal R_{\rm w}$ frame, see ~\cite{pap1} for a detailed discussion. Nevertheless, in order to assess the contribution of such electric fields to the dynamics of the background plasma, we have recorded the trajectories of a large set of particles, and measured  the various contributions to the electromagnetic forces that they suffer in several regions of the precursor. These simulations indicate that the the longitudinal electric forces are dominated by the transverse electromagnetic forces in the near precursor, where most of the heating takes place, but prevail in the far precursor. This may be expected on the grounds that the ``Weibel frame'' is itself well defined in the near precursor, but not necessarily so in the far precursor because of the stronger contribution of oblique modes. 

We also note that the dynamics of a pair shock differs significantly from that of electrons in an electron-ion shock, in which longitudinal electric fields are observed to play an important role~\cite{Kumar_2015}. In the present case, electrons and positrons carry the same inertia and account for the totality of the background plasma. Therefore, in the shock frame, the shock transition can be described in terms of momentum transfer between the longitudinal and the transverse directions through angular relaxation, instead of heating, even though in the noninertial $\mathcal R_{\rm w}$ frame, particle energization does occur. In our model, this angular relaxation occurs gradually over the near precursor, as the plasma slowly relaxes in a turbulence frame, which itself decelerates because of the growing influence of the suprathermal particles. Accordingly, the hydrodynamic shock jump conditions indicate that the energy per particle is conserved between far upstream and far downstream~\cite{1976PhFl...19.1130B}. By contrast, the electrons of the background plasma in an electron-ion shock form an open system that draws energy from the ion reservoir, so that electron energization truly occurs in the shock frame.

One may tentatively extrapolate the present model to the case of electron-ion shocks as follows. In these shocks, the inertia carrier are ions and therefore, in a first approximation, their dynamics is likely similar to that of the pairs in a pair shock. Yet the disparity in inertia between ions and electrons should translate into different scattering frequencies, which would break the equivalence between their trajectories in the effective gravity field associated with  the slowdown of $\mathcal R_{\rm w}$. Hence, one may then expect a coherent longitudinal $\delta E_x$ field to arise and to contribute strongly to electron heating, at the expense of the ions. We note that such a field could not have been seen in the simulations of Ref.~\cite{Kumar_2015}, because a symmetric counterstreaming configuration was then adopted, in which there is no net $\beta_{\rm w}$, hence not net deceleration. Such study is left for further work.

\section{Summary -- conclusions}\label{sec:conc}
The present paper is the second in a series of papers that address the microphysics of a relativistic unmagnetized collisionless pair shock. In Paper~I~\cite{pap1}, it was argued that the filamentation instability, which represents the leading microinstability in the precursor of such collisionless shock waves leads to the generation of a microturbulence of an essentially magnetic nature in a particular reference frame noted $\mathcal R_{\rm w}$, which moves at subrelativistic velocities relative to the background (unshocked) plasma. The present paper makes use of this result to address the issue of deceleration and heating of the background plasma, from its asymptotic values $u_\infty$ ($4-$velocity, $\vert u_\infty\vert\,\gg\,1$) and $T_\infty$ (temperature, $T_\infty\,\ll\,m$) outside the precursor, to the downstream values that are expected to match the shock-crossing values obtained in a fluid limit.

The $\mathcal R_{\rm w}$ frame decelerates in the precursor, because of the presence of suprathermal particles with an $x-$dependent profile. As discussed in Paper~I, indeed, the relative velocity $\beta_{\rm p\vert w}$ of the background plasma in the $\mathcal R_{\rm w}$ is essentially controlled by $\xi_{\rm b}$, which represents the pressure of these suprathermal particles in units of the incoming asymptotic momentum flux $u_\infty^2 n_\infty m$. As background plasma particles cross the precursor, they encounter a finite (and, actually, steadily growing) $\xi_{\rm b}$, which implies an everywhere nonvanishing velocity $\vert\beta_{\rm p\vert w}\vert$. At a fixed background plasma velocity $\beta_{\rm p}$ in the shock frame, this implies a smaller $\vert\beta_{\rm w}\vert$ velocity of $\mathcal R_{\rm w}$ in this shock frame, hence a deceleration of the $\mathcal R_{\rm w}$ frame. Because of the finite value of $\nu_{\vert\rm w}$, the background plasma particles tend to relax towards isotropy at all times in the $\mathcal R_{\rm w}$. This interplay between the background plasma and the microturbulence in the presence of the $\xi_{\rm b}$ profile leads to the deceleration of $\mathcal R_{\rm w}$ and the background plasma in the shock frame. 

As argued in this paper, the above expresses how momentum is transferred from the suprathermal  beam to the background plasma via the microturbulence. In order to obtain the deceleration $u_{\rm p}(x)$, we have used a fluid model which expresses the conservation of energy and momentum between suprathermal particles and the background plasma in the shock precursor. We obtain the law $\left\vert u_{\rm p}(x)\right\vert\,\approx\,\xi_{\rm b}(x)^{-1/2}$ in the deceleration region where $\vert u_{\rm p}\vert\,\lesssim\,\vert u_\infty\vert$, or equivalently $\xi_{\rm b}u_\infty^2\,\gtrsim\,1$. We have compared this prediction to the results of PIC simulations for $\gamma_\infty\,=\,10$ and $\gamma_\infty\,=\,100$, and obtained a satisfactory agreement with the same numerical prefactor of the order of unity in both cases. This law also explains why the typical energy density of suprathermal particles at injection is $\xi_{\rm b}\,\sim\,0.1$: it is at such values that the background plasma slows down to subrelativistic velocities, therefore that the shock transition occurs.

In our description, unmagnetized collisionless shocks are akin to \emph{cosmic-ray-mediated shocks} in which the suprathermal particles transfer momentum to the background plasma through their scattering off the turbulence, which is itself carried by the background plasma~\cite{1981ApJ...248..344D,1987PhR...154....1B}. We show indeed that, within a fluid picture, the growth of the microturbulence plays a negligible role in slowing down the background plasma. Hence, our model departs from the common view that the shock transition is mediated by the buildup of an electromagnetic barrier sitting on the shock front. In the present model, the microturbulence can be described at the fluid level as an MHD-like turbulence carried by the background plasma over most of the precursor, because it is essentially magnetic in the $\mathcal R_{\rm w}$ frame, which nearly coincides with the proper frame of the background plasma. Our model then predicts that the strength parameter $\epsilon_B$, which represents the microturbulence energy density in units of the incoming momentum flux $u_\infty^2 n_\infty m$, must increase as $1/\beta_{\rm w}^2$ through the compression of the transverse magnetic field lines. And indeed, this law is found to reproduce nicely the peak of the $\epsilon_B$ profile seen in PIC simulations at the shock transition. This peak thus does not correspond to a suddenly growing microinstability, but rather to the compression of magnetic field lines as $\mathcal{R}_{\rm w}$ becomes subrelativistic relative to the shock front.

At the kinetic level, the microturbulence plays of course an important role, through pitch-angle scattering, in slowing down and heating the background plasma. To describe this process, we have written a kinetic equation for the distribution of the background plasma, with the spatial variable $x$ along the shock normal expressed in the $\mathcal R_{\rm s}$ shock frame where a stationary solution is sought, and with momentum variables expressed in the $\mathcal R_{\rm w}$ frame. The magnetic nature of the microturbulence in this $\mathcal R_{\rm w}$ frame allows us to express the effective collision operator as pure pitch-angle diffusion, characterized by an average scattering frequency $\nu_{\vert\rm w}$. As the background plasma particles are effectively magnetized in the microturbulence, we treat $\nu_{\vert\rm w}$ as a parameter. 

Here, a crucial point is that the $\mathcal R_{\rm w}$ frame is locally but not globally inertial. Therefore, an inertial correction appears in the kinetic equation for the distribution function, which takes the form of an effective gravity ${\rm d}u_{\rm w}/{\rm d}x$ associated with the slowdown of $\mathcal R_{\rm w}$. The background plasma particles then experience the analog of collisionless Joule heating, in which the effective gravity plays the role of the driving electric field, while their scattering off the microturbulence substitutes itself for collisions. At the microscopic level, particles are heated in $\mathcal R_{\rm w}$ through some form of stochastic shear acceleration. 

In order to follow the plasma heating over all the precursor and to make detailed comparisons with PIC simulations, we have developed a Monte Carlo code, which advances particles subject to random pitch-angle diffusion at a rate $\nu_{\vert\rm w}$ in a noninertial frame $\mathcal R_{\rm w}$. We find that, for a same value $\nu_{\vert\rm w}\,\sim\,0.01\omega_{\rm p}$ and a deceleration profile $u_{\rm w}(x)$ borrowed from our theoretical modeling in Paper~I~\cite{pap1}, it is possible to reproduce satisfactorily the trajectory of heating and deceleration ($u_{\rm p}$ {\it vs} $T_{\rm p}$) observed in our two reference PIC simulations with $\gamma_\infty\,=\,17$ and $\gamma_\infty\,=\,173$. We show that this value of the effective scattering frequency is not only theoretically motivated for background plasma particles interacting with magnetized Weibel filaments, but is also supported by direct measurements from our PIC simulations.

Our analytical model is thus able to describe the main features of the deceleration and heating of the background pair plasma as it flows across the precursor of a relativistic collisionless shock front. In subsequent papers of this series, we will discuss the kinetics of the suprathermal beam particles and the microturbulence dynamics in the shock precursor.

\begin{acknowledgments} 
We acknowledge financial support from the Programme National Hautes \'Energies (PNHE) of the C.N.R.S., the ANR-14-CE33-0019 MACH project and the ILP LABEX (under reference ANR-10-LABX-63) as part of the Idex SUPER, and which is  financed by French state funds managed by the ANR within the "Investissements d'Avenir" program under reference ANR-11-IDEX-0004-02. This work was granted access to the HPC resources of TGCC/CCRT under the allocation 2018-A0030407666 made by GENCI. We acknowledge PRACE for awarding us access to resource Joliot Curie-SKL based in France at TGCC Center.
\end{acknowledgments}

\appendix

\section{Effective Fokker-Planck dynamics in a noninertial frame}\label{app:fp}
The full relativistic transport equation in the diffusion approximation can be found in detailed form in Refs.~\cite{1985ApJ...296..319W,*1989ApJ...340.1112W, 2018MNRAS.479.1747A,*2018MNRAS.479.1771A}. For self-consistency, we borrow these methods to derive here our simplified one-dimensional transport equation in the microturbulence. The first step is to write the Vlasov equation in its  relativistic formulation, as
\begin{equation}
  \frac{{\rm d}x^\mu}{{\rm d}\tau}\frac{\partial}{\partial x^\mu}f
  + \frac{{\rm d}p^i}{{\rm d}\tau}\frac{\partial}{\partial p^i}f\,=\,0\,,
\end{equation}
where $\tau$ denotes proper time, and the distribution function already incorporates the mass-shell condition. It is thus regarded as a function of $\mathbf p$, not $p^0$. It is normalized as usual, with the four-current density defined as
\begin{equation}
  j^\mu\,=\,\int\frac{{\rm d^3}p}{p^0}\,p^\mu\,f\,.
\end{equation}

We now assume that the turbulence is purely magnetic in a frame moving at $\boldsymbol{\beta_{\rm w}}$ with Lorentz factor $\gamma_{\rm w}$ and we rewrite the above Vlasov system in a mixed coordinate system, with the spatial coordinates in the shock frame $\mathcal R_{\rm s}$, and the momenta in the (wave) Weibel frame $\mathcal R_{\rm w}$. The instantaneous or local Lorentz transform from the shock frame to the Weibel frame is characterized by the tetrad ${e^a}_\alpha$ and its inverse ${{e^\star}^\alpha}_a$:
\begin{equation}
  {{p_{\vert\rm w}}}^a\,\equiv\,{e^a}_\alpha\, p^\alpha,\quad
  p^\alpha\,\equiv\,{{e^\star}^\alpha}_a\, {{p_{\vert\rm w}}}^a\,.
\end{equation}
Primed quantities and latin letters $a,b,c$ are associated to the locally (but not globally) inertial frame $\mathcal R_{\rm w}$, while unprimed symbols and greek indices are associated to the (globally inertial) lab frame $\mathcal R_{\rm s}$.

In the mixed coordinate system, the Vlasov equation reads
\begin{equation}
   {{p_{\vert\rm w}}}^a {{e^\star}^\mu}_a\frac{\partial}{\partial x^\mu}f
  + m\frac{{\rm d}{{p_{\vert\rm w}}}^i}{{\rm d}\tau}\frac{\partial}{\partial {{p_{\vert\rm w}}}^i}f\,=\,0\,,
\end{equation}
with
\begin{equation}
  m\frac{{\rm d}{{p_{\vert\rm w}}}^i}{{\rm d}\tau}\,=\, q\, {{F}^i}_a\, {{p_{\vert\rm w}}}^a - {\Gamma}^i_{ab}\,{{p_{\vert\rm w}}}^a{{p_{\vert\rm w}}}^b\,,
\end{equation}
where $m$ denotes the mass of the particles, ${F}^{ab}$ represents the total (coherent + turbulent) electromagnetic field strength tensor in the $\mathcal R_{\rm w}$ frame, indices $i,j, ...$ run over spatial values $1-3$, while $a,b,c...$ run over all four indices. The connection ${\Gamma}^i_{ab}$ accounts for the inertial terms in the comoving wave frame; it is expressed as
\begin{equation}
  \Gamma^a_{bc}\,\equiv\, - {{e^\star}^\beta}_b {{e^\star}^\gamma}_c \frac{\partial}{\partial x^\gamma}{e^a}_{\beta}\,.
\end{equation}
It is not symmetric in its two lower indices, because the tetrad frame is not a coordinate basis. For our particular problem, the non-zero components of the tetrad and its inverse are
\begin{eqnarray}
  {e^t}_t&\,=\,&{e^x}_x\,=\,{{e^\star}^t}_t\,=\,{{e^\star}^x}_x\,=\,\gamma_{\rm w},\nonumber\\
  {e^t}_x&\,=\,&{e^x}_t\,=\,-{{e^\star}^t}_x\,=\,-{{e^\star}^x}_t\,=\,-\gamma_{\rm w}\beta_{\rm w},\nonumber\\
  {e^y}_y&\,=\,&{e^z}_z\,=\,{{e^\star}^y}_y\,=\,{{e^\star}^z}_z\,=\,1\,,
  \label{eq:etet}
\end{eqnarray}
so that the only non-zero components of the connection are
\begin{eqnarray}
  {\Gamma}^t_{xt}&\,=\,&{\Gamma}^x_{tt}\,=\,\frac{1}{\beta_{\rm w}}\partial_t\gamma_{\rm w} + \partial_x\gamma_{\rm w}\,,\nonumber\\
  {\Gamma}^t_{xx}&\,=\,&{\Gamma}^x_{tx}\,=\,\partial_t\gamma_{\rm w} + \frac{1}{\beta_{\rm w}}\partial_x\gamma_{\rm w}\,.
\end{eqnarray}

We further approximate $f\,\simeq\,\langle f\rangle + \delta f$ with $\delta f$ the fluctuating part of $f$. The general Vlasov equation then has a formal solution~\cite{1966PhFl....9.1773D,1969PhFl...12.1045W}:
\begin{align}
  &{{p_{\vert\rm w}}}^a {{e^\star}^\mu}_a\frac{\partial}{\partial x^\mu}\langle f\rangle +
  \left\langle\frac{{\rm d}{{p_{\vert\rm w}}}^i}{{\rm d}\tau}\frac{\partial}{\partial
    {{p_{\vert\rm w}}}^i}\right\rangle\langle f\rangle\,=\,\nonumber\\
&\quad q^2\int {\rm d}\tau\,
  \left\langle {{\delta F}^i}_a {{p_{\vert\rm w}}}^a \frac{\partial}{\partial {{p_{\vert\rm w}}}^i}\,G(t;\,\tau)\,{{\delta F}^j}_b{{p_{\vert\rm w}}}^b
  \frac{\partial}{\partial {{p_{\vert\rm w}}}^j}\right\rangle\langle f\rangle\,,
  \label{eq:quasil}
\end{align}
where $G(t;\tau)$ formally represents the propagator connecting phase space values at time $\tau$ to phase space values at time $t$. In standard quasilinear theory, one approximates the r.h.s. of the above equation to lowest order in the powers of the stochastic force, which amounts to use the unperturbed trajectory in the propagator.

Here we do not make this approximation, which would fail badly in the case of the background plasma. We first note that our assumption of a magneto-static turbulence in the $\mathcal R_{\rm w}$ frame implies that the scattering operator on the r.h.s. of Eq.~(\ref{eq:quasil}) must be purely transverse to $\boldsymbol{{p_{\vert\rm w}}}$. In spherical coordinates, writing ${\mu_{\vert\rm w}}$ the cosine of the angle of $\boldsymbol{{p_{\vert\rm w}}}$ with $\boldsymbol{x}$ and averaging over the azimuthal angle $\phi$, this scattering operator and the resulting equation take the general form:
\begin{align}
 & \gamma_{\rm w}\left({{p_{\vert\rm w}^t}}+\beta_{\rm w} {{p_{\vert\rm w}^x}}\right)\partial_t\langle f\rangle
  + \gamma_{\rm w}\left(\beta_{\rm w} {{p_{\vert\rm w}^t}}+{{p_{\vert\rm w}^x}}\right)\partial_x\langle f\rangle\nonumber\\
&  - \left(\partial_t\gamma_{\rm w}+\frac{\partial_x \gamma_{\rm w}}{\beta_{\rm w}}\right){{p_{\vert\rm w}^t}}{{p_{\vert\rm w}^x}}\partial_{{{p_{\vert\rm w}^x}}}
  \langle f\rangle\nonumber\\
&  - \left(\frac{\partial_t\gamma_{\rm w}}{\beta_{\rm w}}+\partial_x \gamma_{\rm w}\right){{p_{\vert\rm w}^t}}^2\partial_{{{p_{\vert\rm w}^x}}}
  \langle f\rangle  
 + qE_x\,{{p_{\vert\rm w}^t}}\partial_{{{p_{\vert\rm w}^x}}}
 \langle f\rangle \,=\, \nonumber\\
& \quad\quad\frac{{{p_{\vert\rm w}^t}}}{2}\partial_{{\mu_{\vert\rm w}}}\left[\nu_{\vert\rm w}\left(1-{\mu_{\vert\rm w}}^2\right)\right]\partial_{{\mu_{\vert\rm w}}}\langle f\rangle\,,
 \label{eq:quasil2}
\end{align}
where $\nu_{\vert\rm w}$ is a scattering frequency, which a priori depends both on
${\mu_{\vert\rm w}}$ and ${p_{\vert\rm w}}$. We do not aim to calculate $\nu_{\vert\rm w}$ from quasilinear theory; we rather treat this scattering frequency as a parameter of the model.

We now assume a steady state regime in the shock rest frame, so that Eq.~(\ref{eq:quasil2}) above further simplies down to our main equation:
\begin{align}
 & \gamma_{\rm w}\left(\beta_{\rm w} {{p_{\vert\rm w}^t}}+{{p_{\vert\rm w}^x}}\right)\partial_x\langle f\rangle
 - \frac{1}{\beta_{\rm w}}\frac{{\rm d}\gamma_{\rm w}}{{\rm d}x}
 \left(\beta_{\rm w} {{p_{\vert\rm w}^t}}+{{p_{\vert\rm w}^x}}\right){{p_{\vert\rm w}^t}}\partial_{{{p_{\vert\rm w}^x}}}
  \langle f\rangle\nonumber\\
&\quad\quad + qE_x\,{{p_{\vert\rm w}^t}}\partial_{{{p_{\vert\rm w}^x}}}
 \langle f\rangle \,=\, \frac{{{p_{\vert\rm w}^t}}}{2}\partial_{{\mu_{\vert\rm w}}}\left[\nu_{\vert\rm w}\left(1-{\mu_{\vert\rm w}}^2\right)\right]\partial_{{\mu_{\vert\rm w}}}\langle f\rangle\,,
 \label{eq:quasil3}
\end{align}
and, of course,
\begin{equation}
  \partial_{{{p_{\vert\rm w}^x}}}\,\equiv\,{\mu_{\vert\rm w}}\partial_{{p_{\vert\rm w}}} + \frac{1-{\mu_{\vert\rm w}}^2}{{p_{\vert\rm w}}}\partial_{{\mu_{\vert\rm w}}}\,.
\end{equation}
The term proportional to ${\rm d}\gamma_{\rm w}/{\rm d}x$ characterizes the effective gravity felt by the background plasma particles in the noninertial frame $\mathcal R {\rm w}$. In the main part of the text, we omit the brackets $\langle \rangle$ and simply write $f$ for the average part of the distribution function. As we are interested in this paper in the dynamics of a pair plasma, we set $E_x\,=\,0$ because of uniform charge neutrality. This electric field is bound to play a key role in heating the electrons in an electron-ion shock; we defer this discussion to a future study.

\bibliographystyle{apsrev4-1}

\bibliography{shock}

\end{document}